\documentclass[a4paper,11pt]{article}

\usepackage[UTF8]{ctex}
\usepackage[a4paper,margin=1.8cm,columnsep=0.7cm]{geometry}
\usepackage{multicol,multirow,graphicx,epstopdf}
\usepackage[percent]{overpic}
\usepackage{amsmath,amsfonts,amssymb,bm,upgreek,mathrsfs,mathcomp}
\usepackage[pagewise,switch,columnwise]{lineno}
\usepackage[compress,nospace]{cite}
\usepackage[dvipsnames]{xcolor}
\usepackage{booktabs}
\usepackage{array}
\usepackage{siunitx}
\usepackage{float}
\usepackage{caption}
\usepackage[colorlinks=true,linkcolor=blue,urlcolor=blue,citecolor=blue]{hyperref}
\renewcommand{\tablename}{Table}
\renewcommand{\figurename}{Figure}

\renewcommand{\arraystretch}{1.1}
\begin{document}
\title{\textbf{Feasibility Study of $e^+e^-\to \eta_cJ/\psi$ Production and Fully-Charmed Tetraquark Searches at STCF}}
\author{%
	Zhuojuan Dong$^{1}$, Xiaorong Zhou$^{2}$, Xiaorui Lv$^{3}$, Zhengyun You$^{1}$,
	Jin Zhang$^{1,*}$ and Cong Geng$^{1,*}$\\
	\small $^{1}$Sun Yat-Sen University, Guangzhou 510275, China\\
	\small $^{2}$University of Science and Technology of China, Hefei 230026, China\\
	\small $^{3}$University of Chinese Academy of Sciences, Beijing 100049, China\\
	\small $^{*}$Corresponding authors: gengc@mail.sysu.edu.cn; zhangjin5@mail.sysu.edu.cn
}
\date{}
\maketitle

\begin{abstract}
The proposed high-luminosity Super Tau-Charm Facility (STCF) offers a clean experimental environment for threshold studies of exotic hadrons. In this Letter, we evaluate the expected significance for the fully-charmed vector tetraquark candidate $T_{4c}$ in the $e^+e^- \to \eta_c J/\psi$ channel at the STCF. Using Monte Carlo simulations of an energy scan from $\sqrt{s}=6.71$ to $6.79~\mathrm{GeV}$ with $100~\mathrm{fb}^{-1}$ per point, we adopt a high-efficiency single-tag (ST) reconstruction of $J/\psi$ leptonic decays as the primary strategy, with a double-tag (DT) reconstruction of $\eta_c \to K^+K^-\pi^0$ retained as an independent cross-check. Because the signal cross section is governed by the still-uncertain dielectron width $\Gamma_{ee}$, we consider three benchmark hypotheses, $\Gamma_{ee}=0.25, 0.5, 1~\mathrm{eV}$. The corresponding expected ST significances are $5.1\,\sigma$, $10.6\,\sigma$, and $20.5\,\sigma$, respectively. These results indicate that the STCF can provide meaningful sensitivity to fully-charmed tetraquark states near threshold.
\end{abstract}

\section{Introduction}
Quantum Chromodynamics (QCD) predicts a rich spectrum of exotic hadrons beyond the conventional quark model. Over the past two decades, experimental milestones--such as the discoveries of the $X(3872)$\cite{PhysRevLett.91.262001}, the charged $Z_c(3900)$\cite{PhysRevLett.110.252001,PhysRevLett.110.252002}, and the hidden-charm pentaquarks $P_c$\cite{PhysRevLett.115.072001}--have revolutionized our understanding of non-perturbative QCD dynamics.
	
	Recently, fully-charmed tetraquarks ($cc\bar{c}\bar{c}$) have emerged as an important platform for investigating multiquark binding mechanisms and heavy-quark dynamics, as they are free from light-quark degrees of freedom. Following the successive observations of multiple fully-charmed candidate structures, including the $X(6900)$, in the di-$J/\psi$ mass spectrum by the LHCb\cite{lhcbcollaborationObservationStructureJ2020a}, ATLAS\cite{ATLAS2023bft}, and CMS Collaborations\cite{CMS2023owd}, the CMS Collaboration further investigated their quantum numbers through a comprehensive angular analysis\cite{CMS2025fpt}. The results indicate that the tensor state ($J^{PC}=2^{++}$) interpretation is significantly favored by the data, whereas the $0^{++}$, $1^{++}$, and several other low-spin scenarios are constrained to varying degrees. Consequently, these observed structures correspond primarily to the tensor members of the multiplet, leaving their theoretically predicted vector ($J^{PC}=1^{--}$) partners completely unestablished. Since the intermediate virtual photon in $e^+e^-$ annihilation strictly constrains direct resonance production to $1^{--}$ states, $e^+e^-$ collisions offer a pristine and highly selective environment to complete the fully-charmed spectrum by searching for these missing vector states.
	
\begin{figure}[!htb] % 无星号，仅占一栏
		\centering
		\includegraphics[width=7.8cm]{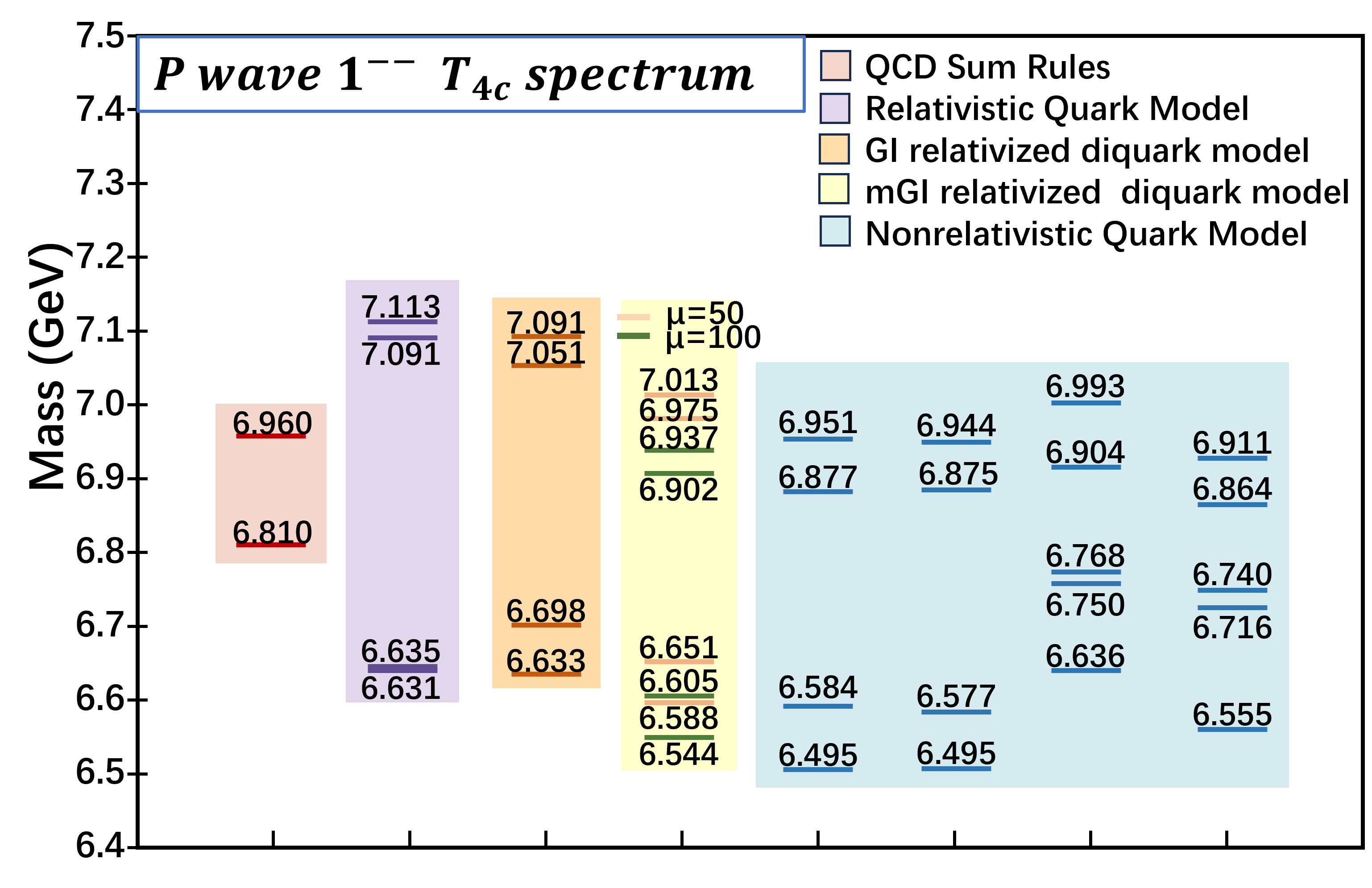} % 宽度缩至单栏宽
		\vskip 2mm
		\caption{Theoretical predictions for the P-wave $1^{--}$ $T_{4c}$ mass spectrum. The shaded columns, from left to right, correspond to QCD sum rules\cite{chenPwaveFullyCharm2024}, the relativistic quark model\cite{faustovFullyheavyTetraquarkSpectroscopy2022}, the Godfrey--Isgur(GI) relativized diquark model\cite{Dong2022sef}, the modified Godfrey--Isgur(mGI) relativized diquark model including color-screening effects\cite{Dong2022sef}, and the nonrelativistic quark model\cite{Dong2022sef,debastianiNonrelativisticModel$ccbarcbarc$2019,liu2020eha,wangHigherFullycharmedTetraquarks2021}. Horizontal bars indicate the predicted mass levels in GeV/$c^2$.}
		\label{fig:theory_mass}
	\end{figure}

	Various theoretical approaches, including QCD sum rules\cite{chenPwaveFullyCharm2024}, relativistic quark models\cite{faustovFullyheavyTetraquarkSpectroscopy2022}, and others\cite{Dong2022sef,debastianiNonrelativisticModel$ccbarcbarc$2019,liu2020eha,wangHigherFullycharmedTetraquarks2021}, predict that P-wave $1^{--}$ fully-charmed tetraquark states lie between 6.5 and 7.1~$\mathrm{GeV}/c^2$ (as illustrated in Fig.~\ref{fig:theory_mass}). Previously, the Belle Collaboration explored this precise threshold region in the $e^+e^-\to\eta_cJ/\psi$ channel via double-charmonium production. However, operating at the $\Upsilon(4S)$ resonance, Belle had to rely on the initial-state radiation (ISR) technique to access this high-mass window\cite{yinSearchDoublecharmoniumState2023}. This ISR approach suffers from a severely suppressed effective luminosity due to the additional electromagnetic coupling and restricted photon phase space. 
	
	To definitively probe this crucial energy regime, the proposed Super Tau-Charm Facility (STCF)\cite{achasovSTCFConceptualDesign2023} is uniquely advantageous. Operating as a symmetric collider up to center-of-mass (c.m.) energy $\sqrt{s}=7.0~\mathrm{GeV}$, the STCF can directly produce these resonances at the c.m. energy without any ISR penalty. Combined with its unprecedented peak luminosity of $\sim0.5\times10^{35}~\mathrm{cm}^{-2}\mathrm{s}^{-1}$, the STCF enables fine energy scans and provides the ultimate precision required to systematically discover and profile these elusive vector exotic resonances.
	
	%-------------------TEXT TEXT TEXT TEXT---------------------
% ======================================================================
% ======================================================================
% Section 2: MC Simulation and STCF Detector
% ======================================================================
\section{STCF Detector and Software Framework}
In this Letter, we present the sensitivity study for the fully-charmed vector tetraquark candidate $T_{4c}$ via the process $e^+e^- \to \eta_c J/\psi$, which relies on simulated data samples representing a realistic data-taking program at the proposed STCF\cite{Lyu:2021tlb}. 
	The simulation is performed with \textsc{OSCAR}, the offline software framework developed for the STCF project. Since the STCF detector design and configuration are still evolving, the \textsc{OSCAR} software is continuously updated accordingly. The framework incorporates the {\sc geant4}-based full detector simulation, including the detector geometry and realistic detector responses used in this analysis, together with Monte Carlo (MC) generators for modeling the underlying physics processes.

%The full detector simulation, which encompasses the geometric description and realistic detector responses of the STCF, is performed within the \textsc{OSCAR} software framework. Furthermore, advanced Monte Carlo (MC) generators are utilized to model the underlying physics processes, optimize event selection, and precisely estimate background contributions.

	The STCF detector\cite{achasovSTCFConceptualDesign2023}, currently under development, is designed as a general-purpose magnetic spectrometer for high-luminosity $e^+e^-$ collisions with nearly $4\pi$ solid-angle coverage. The detector geometry used in this analysis corresponds to the configuration implemented in the \textsc{OSCAR} software at the time of the study, rather than the final STCF detector design, which is still under optimization. In this simulation setup, the sensitive detector systems relevant to the present analysis include the inner tracking system, main drift chamber, particle-identification detectors, electromagnetic calorimeter, and muon detector. These subsystems provide charged-particle tracking, lepton and hadron identification, photon reconstruction, and muon identification, which are the essential detector responses used in the ST and DT analyses.

%In addition, the generic inclusive hadronic backgrounds are modeled with \textsc{Pythia}. 
The \textsc{OSCAR} software is developed to support comprehensive offline data processing, encompassing detector simulation, event reconstruction, and physics analysis\cite{Ai:2024yqx,Li:2024tuy,Li:2024isl,Huang:2023kog}. Designed around the lightweight \textsc{SNiPER} framework, \textsc{OSCAR} integrates fundamental high-energy physics tools such as \textsc{Geant4}\cite{Ivanchenko:2003xp} and \textsc{ROOT}. Additionally, it incorporates state-of-the-art software packages including the Detector Description Toolkit (\textsc{DD4hep}) for geometry management, \textsc{podio} for plain-old-data input/output, and Intel Threading Building Blocks (\textsc{TBB}) for parallel data processing\cite{Shi:2025qmb}. Within the event reconstruction modules, \textsc{OSCAR} deploys sophisticated algorithms to optimize detector performance. Specifically, global track finding is performed based on the Hough transform, assisted by a Graph Neural Network (GNN) for MDC noise filtering\cite{Zhou:2024tio,Jia:2025ufk}. Furthermore, particle identification is driven by advanced machine learning techniques; the GlobalPID package combines information from the tracking system, Cherenkov detectors, EMC, and MUD, and its BDT-based implementation uses 45 detector and track features to provide lepton-identification efficiencies above $90\%$ in the studied momentum range\cite{Zhai:2025qke}. Utilizing this robust architecture, the \textsc{OSCAR} framework performs a realistic full detector simulation that accounts for charged-particle tracking efficiency, momentum resolution, PID responses, and the intrinsic beam-energy spread.

%The MC generation is configured according to the physics category. The \textsc{ConExc} generator is used for the exclusive signal process, the non-resonant $e^+e^- \to \eta_c J/\psi$ continuum, and the dominant exclusive $J/\psi$-associated backgrounds ($J/\psi K^+K^-$, $J/\psi p\bar{p}$, and $J/\psi\pi^+\pi^-$). Following the \textsc{ConExc} prescription for $e^+e^-$ scan experiments, the generation uses input Born cross sections and incorporates initial-state radiation (ISR) effects together with vacuum-polarization corrections. Final-state radiation (FSR) is handled by \textsc{Photos}. Generic inclusive hadronic backgrounds are modeled with \textsc{Pythia}.

MC samples are generated within the OSCAR framework to form the pseudo-data, determine detection efficiencies, and estimate backgrounds. The \textsc{ConExc} generator is used for the signal process $e^+e^- \to T_{4c} \to \eta_c J/\psi$, non-resonant continuum process $e^+e^- \to \eta_c J/\psi$, and backgrounds $e^+e^-\to h^+h^-J/\psi~(h=\pi,K,p)$, incorporating the ISR effects together with vacuum-polarization corrections and final-state radiation handled by \textsc{Photos}\cite{Richter-Was:1992hxq}. The $T_{4c}$ resonance is described by the Breit-Wigner formula given in Eq.~(2).

% ======================================================================
% Section 3: Pseudo-data
% ======================================================================
\section{Methodology and Pseudo-data}
The search sensitivity for the $T_{4c}$ signal is studied through the $\Gamma_{ee}$-dependent Born cross section of the production process $e^+e^-\to \eta_c J/\psi$, which is defined as
\begin{equation} \label{eq:born}
	\sigma_{\rm Born} =
	\frac{N^{ee}_{\rm obs}+N^{\mu\mu}_{\rm obs}}
	{\mathcal{L} f_{\rm ISR} f_{\rm vac} 
		(\varepsilon_{ee}\mathcal{B}_{ee}+\varepsilon_{\mu\mu}\mathcal{B}_{\mu\mu})\mathcal{B}_{\eta_c} }.
\end{equation}
The $N^{ee}_{\rm obs}$ and $N^{\mu\mu}_{\rm obs}$ represent the number of events for the $\eta_c J/\psi$ candidates by analyzing the pseudo-data samples for decay modes $J/\psi\to e^+ e^-$ and $\mu^+ \mu^-$, respectively. The $\mathcal{L}$ labels the integrated luminosity of these pseudo-data samples for each energy point. The $f_{\rm ISR}$ and $f_{\rm vac}$ denote the ISR and vacuum-polarization correction factors. The $\varepsilon_{ee,\mu\mu}$ are the channel-dependent detection efficiencies, and $\mathcal{B}_{ee,\mu\mu}$ are the corresponding $J/\psi \to \ell^+\ell^-$ branching fractions. The $\mathcal{B}_{\eta_c}$ takes into account the branching fraction of the $\eta_c$ decay.

Pseudo-data samples are generated at 17 c.m. energy points from $\sqrt{s}=6.710$ to $6.790~\mathrm{GeV}$ to study the search sensitivity, assuming an integrated luminosity of $100~\mathrm{fb}^{-1}$ per point, which includes the signal process, the dominant background processes, and generic inclusive hadron production.
%
%continuum production of  $e^+e^- \to \eta_c J/\psi$, backgrounds $e^+e^-\to\pi^+\pi^-J/\psi$, $K^+K^-J/\psi$, and $p\bar{p}J/\psi$, and generic inclusive hadron production.
%
%according to respective expected cross sections as the Table~\ref{tab:pseudodata_inputs}. 
%
%
%The background within the selected signal window is well-controlled. Potential contamination from generic inclusive hadronic continuum processes ($e^+e^- \to q\bar{q}$, $q=u,d,s$) is evaluated using a dedicated MC sample of $2 \times 10^6$ events. After event selection, no events survive in the signal region, indicating this background is negligible.
%
The dominant backgrounds consist of non-resonant continuum production of $e^+e^- \to \eta_c J/\psi$ and non-peaking light-hadron production processes accompanied by a $J/\psi$ meson, namely $e^+e^-\to h^+h^-J/\psi~(h=\pi,K,p)$. The cross sections of individual processes are summarized in Table~\ref{tab:pseudodata_inputs}.

%two main components: non-peaking exclusive hadronic processes accompanied by a $J/\psi$ meson, and the irreducible non-resonant continuum production of $e^+e^- \to \eta_c J/\psi$. Based on ISR measurements from Belle~II, three major exclusive channels are identified: $e^+e^-\to\pi^+\pi^-J/\psi$, $K^+K^-J/\psi$, and $p\bar{p}J/\psi$. Dedicated exclusive MC studies confirm that these channels produce smooth distributions in the signal region, preventing the creation of fake resonant structures.

The Born cross section for the signal process $e^+e^- \to T_{4c} \to \eta_c J/\psi$ is calculated using the Breit--Wigner formula\cite{wang2009resonanceparametermeasurementluminosity}:
\begin{equation}
	\sigma_{\rm sig}(s) = \frac{12\pi \, \Gamma_{ee}\,\Gamma_{\eta_c J/\psi}}{(s-m_{T_{4c}}^{2})^{2}+m_{T_{4c}}^{2}\Gamma_{T_{4c}}^{2}}.
	\label{eq:breit-wigner}
\end{equation}
Guided by recent theoretical predictions, the $T_{4c}$ resonance mass and partial width are set to $m_{T_{4c}} = 6.750~\mathrm{GeV}/c^{2}$ and $\Gamma_{\eta_c J/\psi} = 0.17~\mathrm{MeV}$, respectively\cite{liu2020eha}, while its total width is set to $\Gamma_{T_{4c}} = 13~\mathrm{MeV}$\cite{QQQQ2016}. Since no direct experimental information is available for the dielectron width ($\Gamma_{ee}$) of fully-charmed vector tetraquarks, we evaluate the cross sections under three representative hypotheses, $\Gamma_{ee}=0.25, 0.5, 1~\mathrm{eV}$, while keeping the continuum and exclusive-background inputs unchanged. The signal line shapes corresponding to these three hypotheses are used to generate the pseudo-data samples and to assess the ST sensitivity.

\begin{figure}[!htb]
	\centering
	\includegraphics[width=7.8cm]{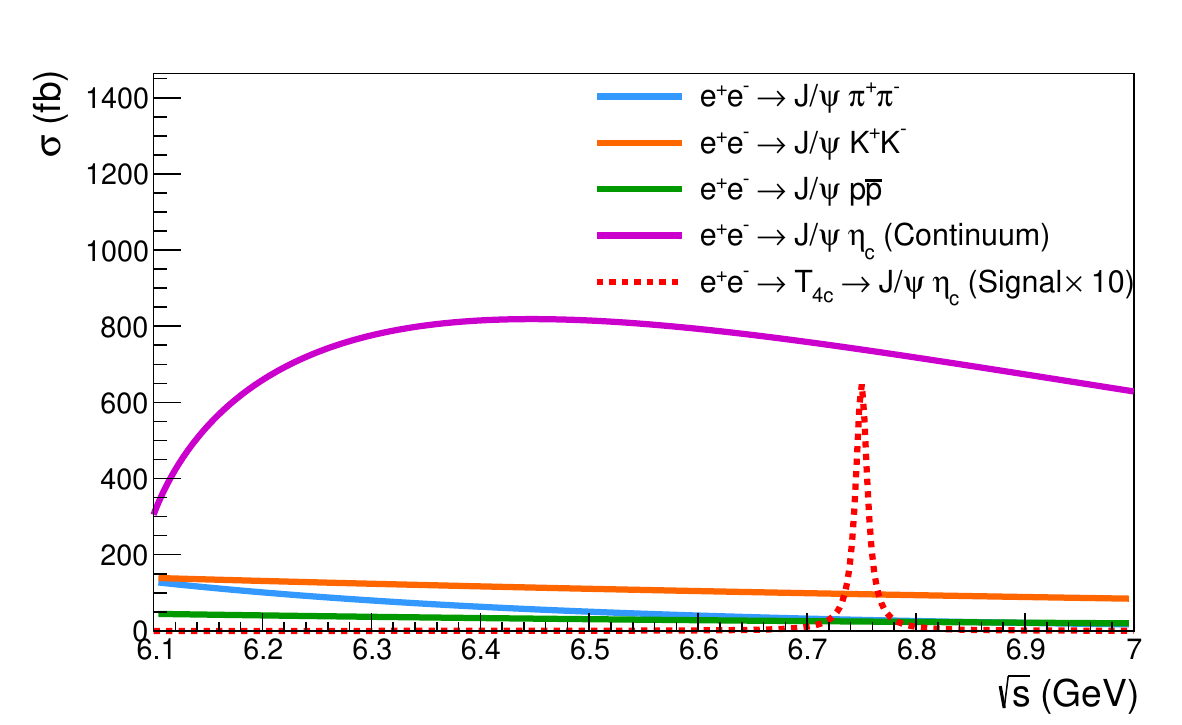}
	\vskip 2mm
	\caption{Input cross-section line shapes used for pseudo-data construction in the STCF scan region. The curves for the three $e^+e^-\to h^+h^-J/\psi~(h=\pi,K,p)$ channels are extrapolated cross-section functions obtained by fitting the Belle~II results. The remaining two curves represent the non-resonant $e^+e^-\to\eta_cJ/\psi$ continuum background and the $e^+e^-\to T_{4c}\to\eta_cJ/\psi$ signal. The $T_{4c}$ signal is shown as a reference line shape and is scaled by a factor of 10 for visibility. All cross sections are shown in fb.}
	\label{fig:lineshape-comparison}
	\vspace{-4mm}
\end{figure}

The non-resonant continuum contribution of $e^+e^- \to \eta_c J/\psi$ is estimated by using the Belle double-charmonium measurement, where the threshold-modulated power-law parametrization $\sigma_{\rm cont}(s) = A \sqrt{2\mu\Delta M}/(s/s_{0,\rm cont})^n$ is used to extrapolate its cross-section line shape to the STCF scan region. Here, $\mu$ is the reduced mass of the charmonium pair, $\Delta M = \sqrt{s} - m_{\eta_c} - m_{J/\psi}$ is the mass excess above threshold, and $\sqrt{s_{0,\rm cont}} \equiv 10.58~\mathrm{GeV}$\cite{yinSearchDoublecharmoniumState2023}. 
To estimate the contributions from the exclusive backgrounds, $e^+e^-\to h^+h^-J/\psi~(h=\pi,K,p)$, the extrapolation is also performed on their Born cross sections by using the Belle~II measurements for these three processes, respectively. 
For each exclusive channel, a power-law decay model $\sigma(\sqrt{s}) = \sigma_0 / (\sqrt{s} / \sqrt{s_{0,\rm excl}})^n$ is used to fit the Belle~II results and obtain the extrapolated cross-section function. The resulting exclusive-background line shapes, together with the non-resonant $e^+e^-\to\eta_cJ/\psi$ continuum background and the $T_{4c}$ signal, are shown in Fig.~\ref{fig:lineshape-comparison}. The $T_{4c}$ signal is shown as a reference line shape and is scaled by a factor of 10 in Fig.~\ref{fig:lineshape-comparison} for visibility.
Here, $\sqrt{s_{0,\rm excl}} = 4.50~\mathrm{GeV}$ is the reference c.m. energy scale for the exclusive-background extrapolation, $\sigma_0$ is the normalization cross section, and $n$ is the suppression index\cite{BelleII2026heb}. All the reference cross-section inputs used for pseudo-data construction are summarized in Table~\ref{tab:pseudodata_inputs}. These values correspond to the scan point, $\sqrt{s}=6.750~\mathrm{GeV}$, where the $T_{4c}$ signal reaches its peak value. The point-by-point ST and DT efficiencies, correction factors, and signal and continuum cross sections used for the updated scan are provided in the Supplementary Material.

\begin{table}[!htb]
	\centering
	\caption{Cross-section inputs at $\sqrt{s}=6.750~\mathrm{GeV}$ for the three $\Gamma_{ee}$ hypotheses.}
	\label{tab:pseudodata_inputs}
	\vskip 2mm
	\small
	\renewcommand{\arraystretch}{1.25}
	\setlength{\tabcolsep}{3.5pt}
	\begin{tabular}{@{}p{3.55cm}ccc@{}} 
		\hline\hline\hline
		Component & \multicolumn{3}{c}{$\sigma~[\mathrm{fb}]$ for $\Gamma_{ee}~[\mathrm{eV}]$} \\
		\cline{2-4}
		& 0.25 & 0.5 & 1.0 \\
		\hline
		$e^+e^-\to T_{4c}\to\eta_cJ/\psi$ & $81.0$ & $162.0$ & $324.0$ \\
		$e^+e^-\to\eta_cJ/\psi~$ & \multicolumn{3}{c}{$739.2$} \\
		$e^+e^-\to\pi^+\pi^-J/\psi$ & \multicolumn{3}{c}{$28.9$} \\
		$e^+e^-\to K^+K^-J/\psi$ & \multicolumn{3}{c}{$63.3$} \\
		$e^+e^-\to p\bar{p}J/\psi$ & \multicolumn{3}{c}{$24.5$} \\
		\hline\hline\hline
	\end{tabular}
\end{table}

%To evaluate the discovery potential, the final pseudo-data samples at each energy point are explicitly constructed by superimposing three categories of simulated events according to their expected cross sections and the designed $\mathcal{L} = 100~\mathrm{fb}^{-1}$ integrated luminosity per point: (i) the fully-charmed tetraquark signal; (ii) the irreducible non-resonant continuum ($e^+e^-\to \eta_c J/\psi$); and (iii) the dominant exclusive hadronic backgrounds. These inputs define the reference pseudo-data sample used for the subsequent ST-dominated sensitivity evaluation.

% ======================================================================
% Section 4: Event Selection
% ======================================================================
\section{Event Selection}
We search for the fully-charmed vector tetraquark candidate in the $e^{+}e^{-}\to T_{4c}\to \eta_{c}J/\psi$ process using two complementary strategies: a high-efficiency single-tag (ST) approach reconstructing only the $J/\psi$ through its leptonic decays $J/\psi \to \ell^+\ell^-$ ($\ell=e,\mu$), and an exclusive double-tag (DT) approach that specifies the final state by fully reconstructing both charmonia through $\eta_c \to K^+K^-\pi^0$ and $J/\psi \to \ell^+\ell^-$, with $\pi^0\to \gamma\gamma$.

Charged tracks detected in the ITK and MDC are required to be within a polar angle range of $|\!\cos\theta| < 0.94$, where $\theta$ is defined with respect to the beam direction. Charged tracks must originate from the interaction point (IP), with distances of closest approach satisfying $|V_r|\le0.1~\mathrm{cm}$ in the plane perpendicular to the beam and $|V_z|\le0.5~\mathrm{cm}$ along the beam direction. In addition, charged tracks are required to have transverse momentum $p_T\ge50~\mathrm{MeV}/c$. Particle identification (PID) is performed using the BDT-based GlobalPID algorithm. Photon candidates are reconstructed from EMC showers with the EMC timing required to be within 10~ns of the event start time.

In the ST approach, the $J/\psi$ candidates are reconstructed through a pair of $e^+e^-$ or $\mu^+\mu^-$, with each lepton required to have momentum greater than $0.9~\mathrm{GeV}/c$. A vertex fit tool is performed to constrain the two leptons to originate from a common vertex, and the resulted fit $\chi^2$ is required to be less than 60. The $J/\psi$ candidates are required to satisfy $3.05 \le M(\ell^{+}\ell^{-}) \le 3.15~\mathrm{GeV}/c^{2}$. The $\eta_c$ signal is inferred from the invariant mass of the system recoiling against the selected $J/\psi$, defined as improved mass variable $M_{\mathrm{rec}}^{\mathrm{corr}}(\ell^{+}\ell^{-}) \equiv M_{\mathrm{rec}}(\ell^{+}\ell^{-}) + M(\ell^{+}\ell^{-}) - m_{J/\psi}$. The variable $M_{\mathrm{rec}}(\ell^{+}\ell^{-})$ denotes the invariant mass recoiling against the lepton pair, the $M(\ell^{+}\ell^{-})$ represents the invariant mass of the lepton pair, and $m_{J/\psi}$ labels the world average value of $J/\psi$ mass from PDG. At the representative energy point $\sqrt{s}=6.750~\mathrm{GeV}$, the detection efficiencies are $64.9\%$ and $79.5\%$ for the electron and muon channels, respectively.

%events must contain at least one positively charged and one negatively charged track. The $J/\psi$ candidate is reconstructed from an oppositely charged lepton pair, with each lepton required to have momentum greater than $0.9~\mathrm{GeV}/c$, and the two leptons are constrained to originate from a common vertex with $\chi^2<60$. The selected $J/\psi$ candidates are required to satisfy $3.05 \le M(\ell^{+}\ell^{-}) \le 3.15~\mathrm{GeV}/c^{2}$. The inclusive $\eta_c$ signal is inferred from the corrected recoil mass spectrum, defined as $M_{\mathrm{rec}}^{\mathrm{corr}} \equiv M_{\mathrm{rec}}(J/\psi) + M(\ell^{+}\ell^{-}) - m(J/\psi)$, where $M_{\mathrm{rec}}(J/\psi) = \sqrt{(p_{e^{+}e^{-}}-p_{J/\psi})^{2}}/c$.

In the DT approach, the $\eta_c$ is reconstructed via the decay mode $K^+K^-\pi^0$ and $\pi^0\to\gamma\gamma$ along with the $J/\psi$ through the lepton pair. The lepton selections are the same as those in the ST approach. The $\pi^0$ candidate is reconstructed with a photon pair within the invariant-mass region (0.1, 0.15)~GeV/$c^2$. To improve the resolution, a kinematic fit is performed by constraining the invariant mass of the photon pair to be the $\pi^0$ mass and requiring the corresponding $\chi^2$ of the fit to be less than 200. The momenta updated by the kinematic fit are used in the further analysis. The final state $K^+K^-\pi^0\ell^+\ell^-$ is constrained to the four-momentum of the initial electron-positron collision with a kinematic fit under energy-momentum conservation, and the resulting $\chi^2$ is required to be less than 200. If there are multiple $\pi^0$ candidates, the combination with the smallest $\chi^2$ value of the kinematic fit is retained for the further analysis. By fully reconstructing the specified $K^+K^-\pi^0\ell^+\ell^-$ final state, the DT strategy provides a more exclusive event definition and an independent cross-check of the ST measurement. At the representative energy point $\sqrt{s}=6.750~\mathrm{GeV}$, the detection efficiencies are $30.85\%$ and $39.79\%$ for the electron and muon modes, respectively.

%events must contain at least two positively charged tracks, two negatively charged tracks, and two good photon candidates. After PID, exactly one positively charged and one negatively charged lepton are required to form the $J/\psi$ candidate, with each lepton required to have momentum greater than $0.9~\mathrm{GeV}/c$ and with the same vertex-fit requirement of $\chi^2<60$ as in the ST analysis. The $\pi^0$ candidate is reconstructed with a photon pair within the invariant-mass region (0.1, 0.15)~GeV/$c^2$; a mass-constrained kinematic fit is then applied, requiring $\chi^2<200$. In addition, exactly one $K^+$ and one $K^-$ candidate are selected by PID, and the $K^+K^-$ pair is required to satisfy a vertex-fit requirement of $\chi^2<200$. The full $K^+K^-\pi^0\ell^+\ell^-$ final state is further subjected to a four-constraint (4C) kinematic fit to the initial $e^+e^-$ four-momentum, with $\chi^2_{\mathrm{4C}}<200$.

%By fully reconstructing both charmonia, the DT strategy suppresses combinatorial backgrounds and provides a high-purity cross-check of the ST measurement. and $40.9\%$ and $34.1\%$ for the corresponding DT modes.
After the above selections, the ST approach extracts the inclusive $\eta_c$ signal from the corrected recoil-mass spectrum, while the DT analysis uses the simultaneous constraints from the reconstructed $J/\psi$ and $\eta_c$ masses to provide an independent validation in a fully specified exclusive final state.

% ======================================================================
% Section 5: Signal Extraction
% ======================================================================
\section{Sensitivity Determination}
\begin{figure}[!htb]
	\centering
	\includegraphics[width=7.8cm]{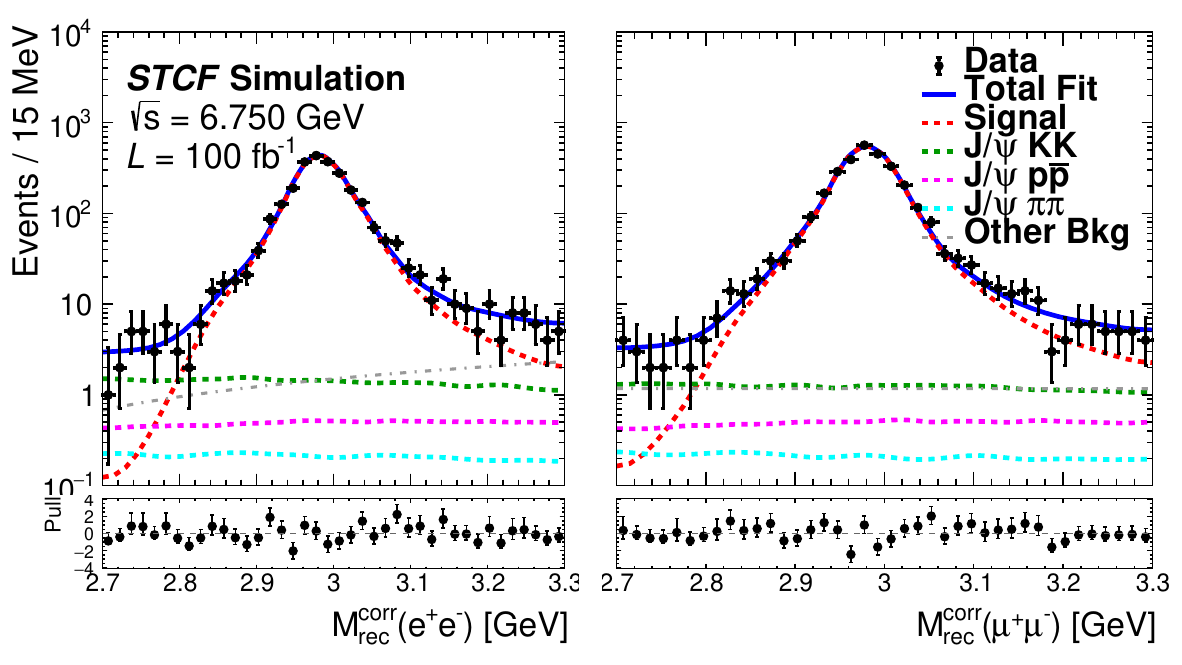}
	\vskip 2mm
\caption{Representative unbinned maximum-likelihood fits to the $M_{\rm rec}^{\rm corr}(\ell^+\ell^-)$ distributions at $\sqrt{s}=6.750~\mathrm{GeV}$. The left and right panels show the $e^+e^-$ and $\mu^+\mu^-$ channels, respectively.}
	\label{fig:yield-fits-st}
	\vspace{-4mm}
\end{figure}

The sensitivity is determined from the Born cross section of the production $e^+e^-\to \eta_c J/\psi$, as defined in Eq.~(\ref{eq:born}).
%The sensitivity is studied based on the Born cross sections of the production $e^+e^-\to \eta_c J/\psi$, which is determined as
%\begin{equation}
%	\sigma_{\rm Born} =
%	\frac{N^{ee}_{\rm fit}+N^{\mu\mu}_{\rm fit}}
%	{\mathcal{L} \cdot f_{\rm ISR} \cdot f_{\rm vacuum} \cdot
%		(\varepsilon_{ee}\mathcal{B}_{ee}+\varepsilon_{\mu\mu}\mathcal{B}_{\mu\mu})}.
%\end{equation}
%The extracted ST signal yields in the electron and muon channels, $N^{ee}_{\rm fit}$ and $N^{\mu\mu}_{\rm fit}$, are summed and converted to Born cross sections via:
%\begin{equation}
%	\sigma_{\rm Born} =
%	\frac{N^{ee}_{\rm fit}+N^{\mu\mu}_{\rm fit}}
%	{\mathcal{L} \cdot f_{\rm ISR} \cdot f_{\rm vacuum} \cdot
%		(\varepsilon_{ee}\mathcal{B}_{ee}+\varepsilon_{\mu\mu}\mathcal{B}_{\mu\mu})}.
%\end{equation}
%where 
%The $\mathcal{L} = 100~\mathrm{fb}^{-1}$ per energy point, $f_{\rm ISR}$ and $f_{\rm vacuum}$ denote the ISR and vacuum-polarization correction factors, $\varepsilon_{ee,\mu\mu}$ are the channel-dependent detection efficiencies, and $\mathcal{B}_{ee,\mu\mu}$ a e the corresponding $J/\psi \to \ell^+\ell^-$ branching fractions. For the DT cross-section determination, the same procedure is used with the additional branching fraction $\mathcal{B}(\eta_c\to K^+K^-\pi^0)$ included in the normalization. Details are provided in the Supplementary Material.
%
In the ST approach, the parameter $\mathcal{B}_{\eta_c}$ is not taken into account. The signal yields $N^{ee}_{\rm obs}$ and $N^{\mu\mu}_{\rm obs}$ are extracted by performing unbinned maximum likelihood fits to the distributions of $M_{\rm rec}^{\rm corr}$ from pseudo-data at each energy point for the $e^+e^-$ and $\mu^+\mu^-$ channels, respectively. The signal shape for the $\eta_c$ is derived from the simulated signal events, and the smooth background under the $\eta_c$ resonance is modeled by a first-order Chebyshev polynomial function. For the $\Gamma_{ee}=0.25~\mathrm{eV}$ hypothesis, representative fits at $\sqrt{s}=6.750~\mathrm{GeV}$ are shown in Fig.~\ref{fig:yield-fits-st}; the left and right panels show the $e^+e^-$ and $\mu^+\mu^-$ channels, respectively. The fitted signal yields are $2531.3\pm53.1$ and $3020.8\pm57.8$ events for the $e^+e^-$ and $\mu^+\mu^-$ channels, respectively. The corresponding summed background contributions are $144.8\pm21.2$ and $125.4\pm21.2$ events, corresponding to $B/S\simeq5.7\%$ and $4.2\%$, respectively.

\begin{figure}[!htb]
\centering
\includegraphics[width=7.8cm]{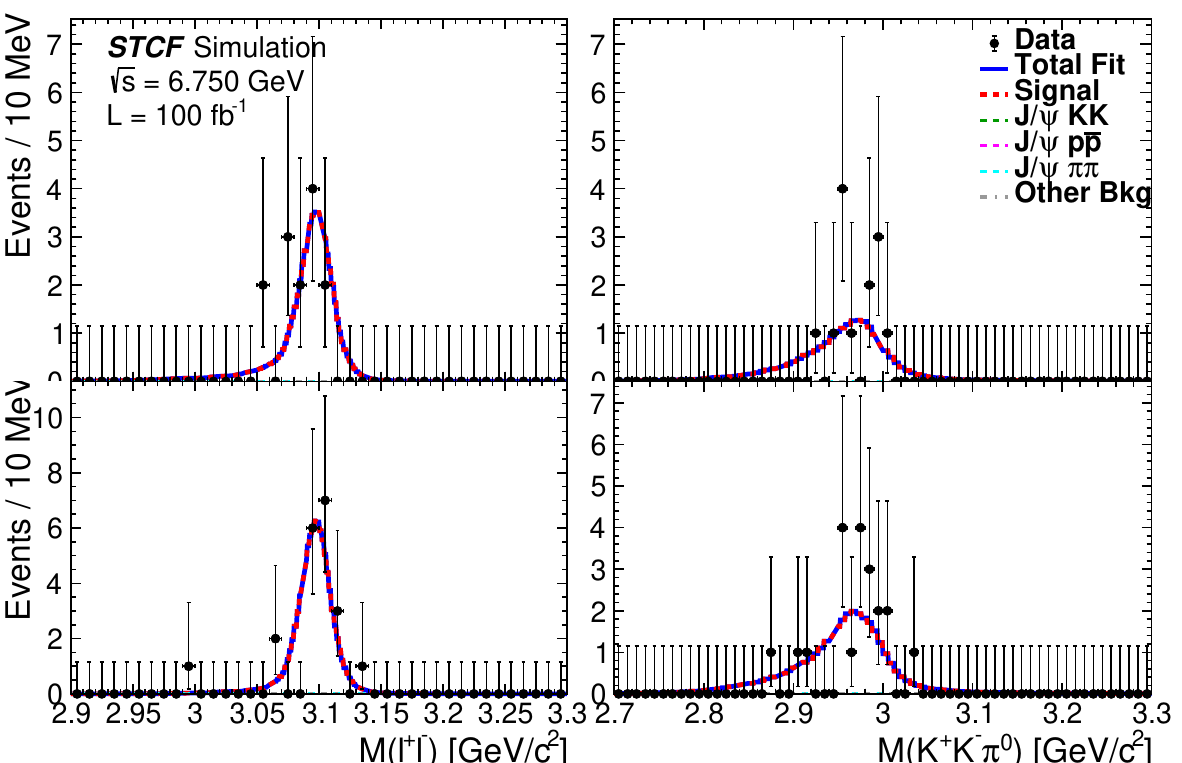}
\vskip 2mm
\caption{Representative projected DT mass fits at $\sqrt{s}=6.750~\mathrm{GeV}$. The left column shows the $M(\ell^+\ell^-)$ distributions and the right column shows the $M(K^+K^-\pi^0)$ distributions; the upper and lower rows correspond to the $e^+e^-$ and $\mu^+\mu^-$ channels, respectively.}
\label{fig:dt-2d-fit-combined}
\vspace{-4mm}
\end{figure}

The DT approach is performed as an independent cross-check with a fully specified exclusive final state but larger statistical uncertainty. A two-dimensional (2D) unbinned maximum-likelihood fit is performed on the distribution of $M(\ell^+\ell^-)$ versus $M(K^+K^-\pi^0)$, where the 2D signal shape is derived from simulated signal MC events. For background components of non-peaking light-hadron production processes accompanied by a $J/\psi$ meson, the corresponding simulated shapes are used, while the remaining smooth background contributions are modeled with first-order Chebyshev polynomial functions, following the treatment used in the ST approach. For the $\Gamma_{ee}=0.25~\mathrm{eV}$ hypothesis, representative fits at $\sqrt{s}=6.750~\mathrm{GeV}$ are shown in Fig.~\ref{fig:dt-2d-fit-combined}; the upper and lower rows show the $e^+e^-$ and $\mu^+\mu^-$ channels, respectively, while the left and right columns show the $M(\ell^+\ell^-)$ and $M(K^+K^-\pi^0)$ projections. The corresponding fitted signal yields are $13.0\pm3.4$ and $19.9\pm4.3$ events for the electron and muon channels, respectively; the fitted background contributions are negligible.
%and corresponding significances are less than those from ST, as shown in Table~\ref{tab:final-significance}.

\begin{figure*}[!t]
	\centering
	\begin{minipage}{5.4cm}
		\centering
		\begin{overpic}[width=5.4cm]{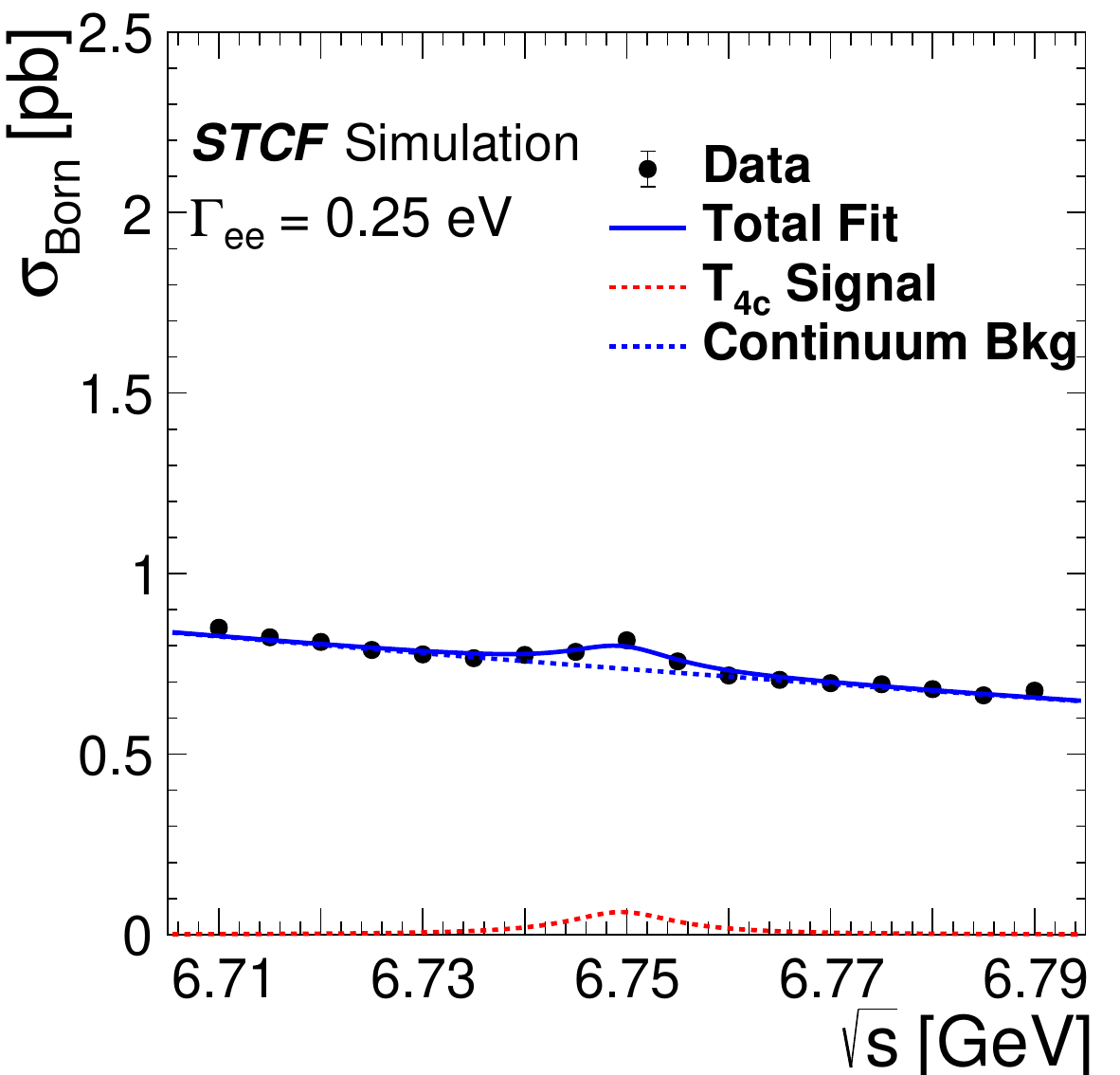}
			\put(74,22){\colorbox{white}{\scriptsize\bfseries (a)}}
		\end{overpic}
		\end{minipage}%
		\hspace{0.6mm}%
	\begin{minipage}{5.4cm}
		\centering
		\begin{overpic}[width=5.4cm]{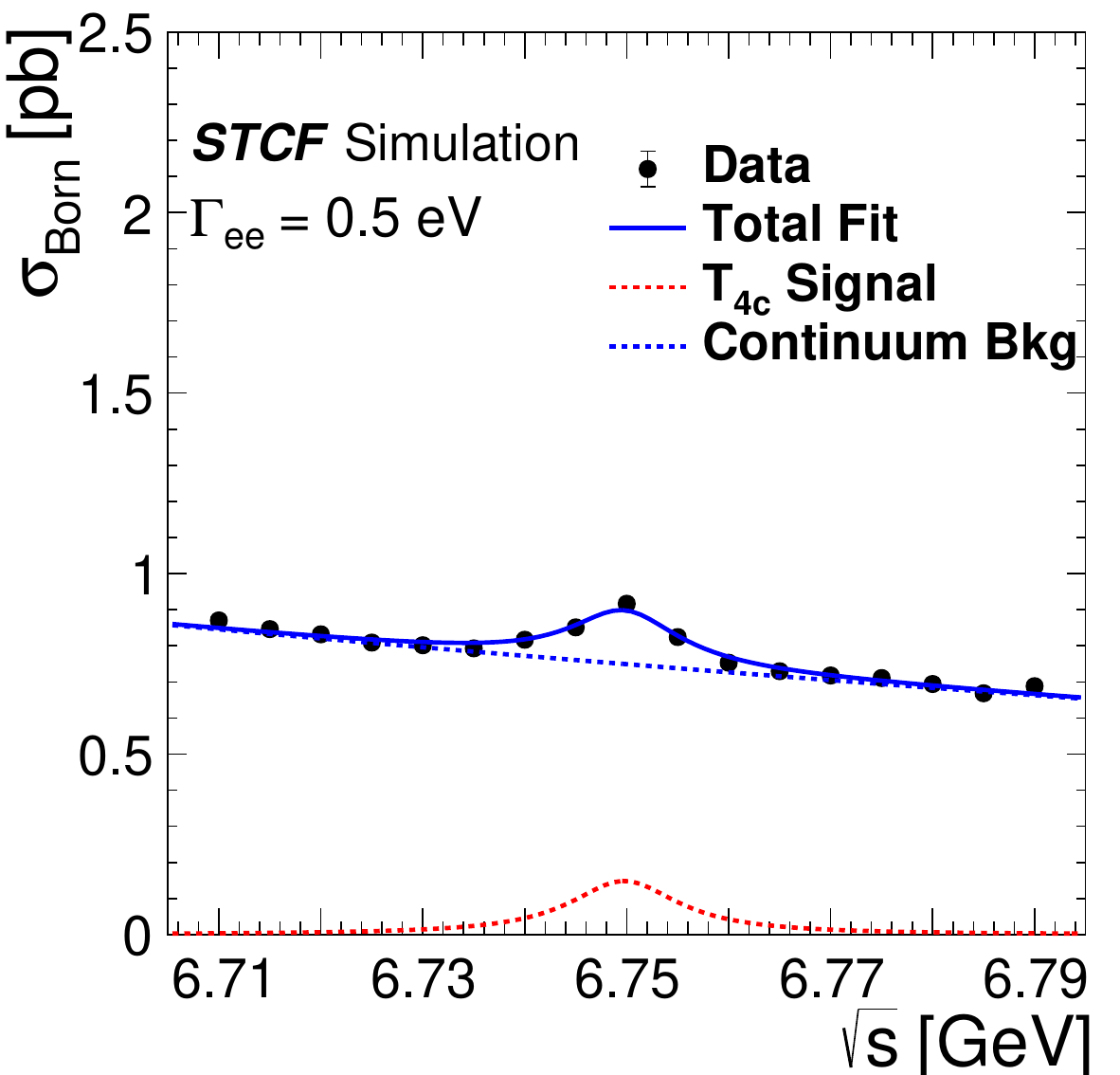}
			\put(74,22){\colorbox{white}{\scriptsize\bfseries (b)}}
		\end{overpic}
		\end{minipage}%
		\hspace{0.6mm}%
	\begin{minipage}{5.4cm}
		\centering
		\begin{overpic}[width=5.4cm]{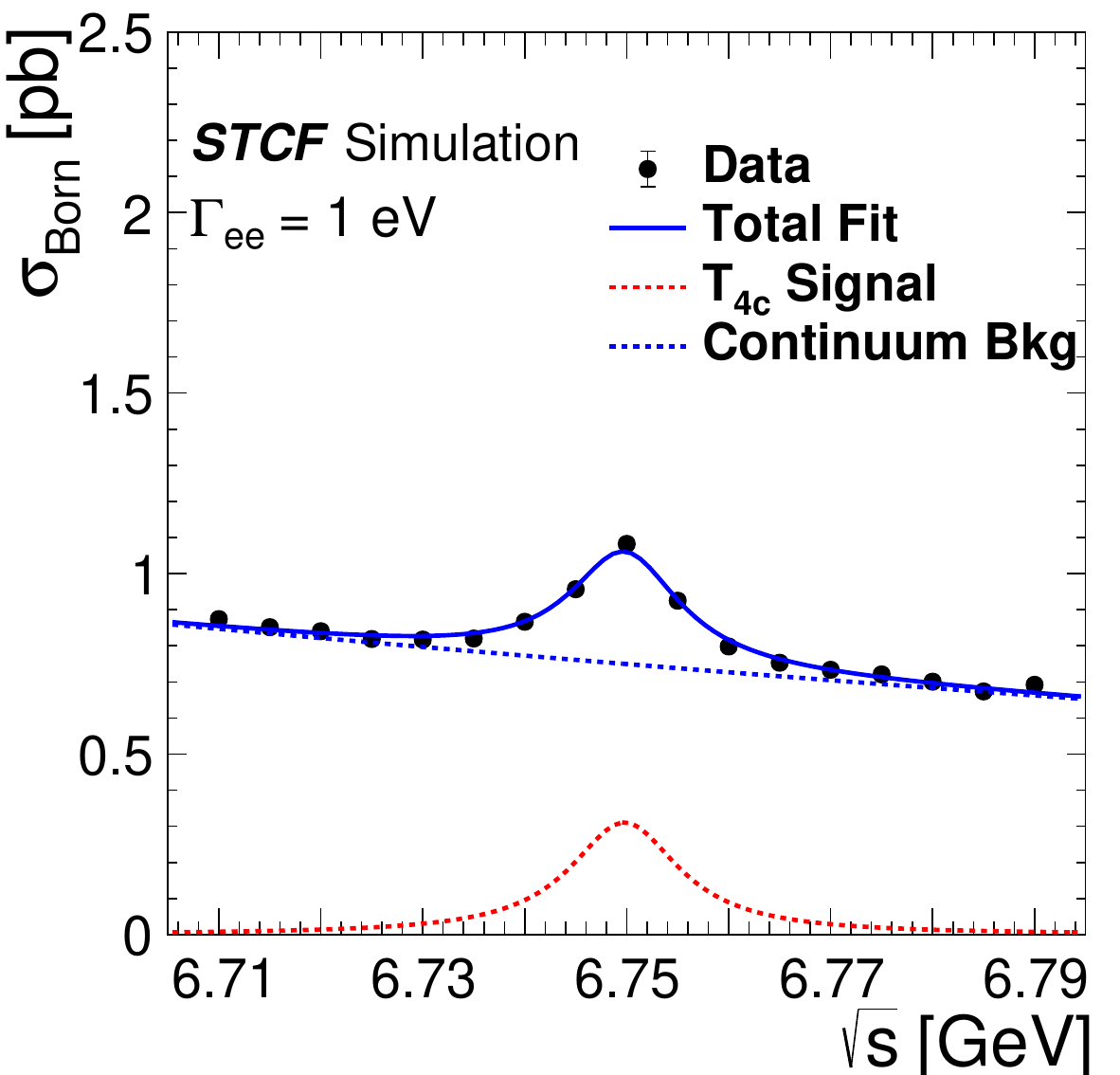}
			\put(74,22){\colorbox{white}{\scriptsize\bfseries (c)}}
		\end{overpic}
		\end{minipage}
		\vskip 2mm
\caption{Single-tag cross-section results for $\Gamma_{ee}=0.25$, $0.5$, and $1~\mathrm{eV}$, shown in panels (a)--(c), respectively. The solid blue curves represent the total curve, the dotted blue curves indicate the continuum component, and the dotted red curves show the $T_{4c}$ signal component. The corresponding DT fits are provided in the Supplementary Material.}
	\label{fig:cross-section-fits-all}
	\vspace{-4mm}
\end{figure*}

Finally, the Born cross sections across the scanned energy points are presented in Fig.~\ref{fig:cross-section-fits-all}. Panels (a)--(c) show the distributions from the ST approach under the $\Gamma_{ee}=0.25$, $0.5$, and $1~\mathrm{eV}$ hypotheses, respectively.
%In both strategies, the fitted signal component appears as a resonance-like enhancement on top of the non-resonant continuum contribution from $e^+e^-\to \eta_c J/\psi$. 
%
To quantify the expected significance, a combined $\chi^2$ fit is performed on the measured cross-section spectrum. The uncertainty assigned to each cross-section point is obtained by adding the statistical uncertainty and the systematic uncertainty in quadrature. The significance is estimated from the change in $\chi^2$ between fits with and without the signal component.
The ST strategy gives expected significances of $5.1\,\sigma$, $10.6\,\sigma$, and $20.5\,\sigma$ for the three $\Gamma_{ee}$ hypotheses, respectively, while the corresponding DT cross-check gives $0.7\,\sigma$, $1.9\,\sigma$, and $3.1\,\sigma$. The DT cross-section fits for the three $\Gamma_{ee}$ hypotheses are provided in the Supplementary Material.

%with the ST strategy as the primary measurement channel. For the ST topology, independent 1D unbinned maximum likelihood fits are performed on the $M_{\rm rec}^{\rm corr}$ spectra at each scan point for the $e^+e^-$ and $\mu^+\mu^-$ channels. The probability density function (PDF) consists of a shape template derived from signal MC and a first-order Chebyshev polynomial background. A representative ST fit is shown in Fig.~\ref{fig:yield-fits-mumu}.

%The background treatment is validated with dedicated MC samples after applying the full event selection. For the generic inclusive hadronic background, a sample of $2 \times 10^6$ events is tested, and no events survive in the signal fitting region. For the exclusive $J/\psi$-associated backgrounds, $5 \times 10^5$ events are simulated for each of the $J/\psi K^+K^-$, $J/\psi p\bar{p}$, and $J/\psi\pi^+\pi^-$ channels. After the selection, their distributions in the signal fitting region are found to be flat and non-peaking, supporting the use of a first-order Chebyshev polynomial to describe the residual background in the ST recoil-mass fit.

%The global discovery significances, incorporating the impact of systematic uncertainties, are summarized in Table~\ref{tab:final-significance}. The ST strategy yields projected significances of $5.1\,\sigma$, $10.6\,\sigma$, and $20.5\,\sigma$ for $\Gamma_{ee}=0.25$, $0.5$, and $1~\mathrm{eV}$, respectively, with the DT topology providing a lower-significance cross-check in each case. These projected results highlight the potential of the proposed STCF to systematically search for and investigate fully-charmed exotic tetraquark states.

\begin{table}[!htb]
	\centering
	\caption{The signal cross sections and projected significances for the three $\Gamma_{ee}$ hypotheses. The reported significances include the impact of systematic uncertainties.}
	\label{tab:final-significance}
	\vskip 2mm
	\small 
	\renewcommand{\arraystretch}{1.3}
	\begin{tabular}{cccc}
		\hline\hline\hline
		$\Gamma_{ee}$ ($\mathrm{eV}$) & \begin{tabular}{c}Signal cross section\\ ($\mathrm{fb}$)\end{tabular} & ST & DT \\
		\hline
		0.25 & 81.0  & $5.1\,\sigma$  & $0.7\,\sigma$ \\
		0.5  & 162.0 & $10.6\,\sigma$ & $1.9\,\sigma$ \\
		1.0  & 324.0 & $20.5\,\sigma$ & $3.1\,\sigma$ \\
		\hline\hline\hline
	\end{tabular}
\end{table}

\section{Systematic Uncertainties}
As a prospective estimate for the STCF, statistics-driven components of the experimental uncertainties are scaled according to the expected luminosity increase relative to BESIII. These estimates are used to assign the systematic uncertainty on each measured cross-section point, which is then combined in quadrature with the corresponding statistical uncertainty in the cross-section fits. The resulting systematic-uncertainty estimate is summarized in Table~\ref{tab:systematics_summary}.

\begin{table}[!htb]
	\centering
	\caption{The estimated systematic uncertainties ($\%$) for the cross-section measurement.}
	\label{tab:systematics_summary}
	\vskip 2mm
	\small
	\renewcommand{\arraystretch}{1.3}
	\begin{tabular}{lcc}
		\hline\hline\hline
		\multirow{2}{*}{Source} & \multicolumn{2}{c}{Uncertainty (\%)} \\
		\cline{2-3}
		& Single-Tag & Double-Tag \\
		\hline
		Tracking efficiency & 0.28 & 0.56 \\
		PID efficiency & 0.28 & 0.56 \\
		Luminosity & 1.0 & 1.0 \\
		$\mathcal{B}(J/\psi \to \ell^+\ell^-)$ (PDG) & 0.55 & 0.55 \\
		$\mathcal{B}(\eta_c \to K^+K^-\pi^0)$ & -- & 13.6 \\
		\hline
		\textbf{Total} & \textbf{1.2} & \textbf{13.7} \\
		\hline\hline\hline
	\end{tabular}
\end{table}

For the ST strategy, the dominant experimental uncertainties arise from tracking and PID efficiencies, luminosity, and the external branching fraction of $J/\psi \to \ell^+\ell^-$. Taking a typical $1\%$ per-track uncertainty at BESIII as a reference, the scaling by the $\sqrt{50}$ statistics enhancement gives an expected uncertainty of roughly $0.14\%$ per track at the STCF. Consequently, for the two-track ST reconstruction, we assign $0.28\%$ for both tracking and PID. Together with the luminosity uncertainty ($1.0\%$) and the external $J/\psi \to \ell^+\ell^-$ branching fraction error ($0.55\%$), the total systematic uncertainty for the ST measurement is estimated to be $1.2\%$. The DT strategy, involving four charged tracks and the external $\eta_c \to K^+K^-\pi^0$ branching fraction, carries a substantially larger total uncertainty of $13.7\%$, reinforcing its role as a cross-check rather than the primary sensitivity driver.

\section{Conclusion}
In this Letter, we present a feasibility study for searching for the fully-charmed vector tetraquark candidate $T_{4c}$ in the $e^{+}e^{-}\to\eta_{c}J/\psi$ channel at the proposed STCF. Utilizing the \textsc{OSCAR} software framework alongside detailed pseudo-data analyses, an energy scan was evaluated in the range of $\sqrt{s}=6.71\text{--}6.79~\mathrm{GeV}$ with an assumed integrated luminosity of $100~\mathrm{fb}^{-1}$ per point.
Our study indicates that the ST strategy, which infers the inclusive $\eta_{c}$ via the recoil mass against the reconstructed $J/\psi$ candidates, is the primary discovery channel because it has substantially higher signal efficiency than the exclusive DT method.
% ($\sim65\%$--$79\%$ versus $\sim34\%$--$41\%$). 
Incorporating the impact of systematic uncertainties, estimated to be approximately $1.2\%$, the ST approach provides projected significances of $5.1\,\sigma$, $10.6\,\sigma$, and $20.5\,\sigma$ for $\Gamma_{ee}=0.25$, $0.5$, and $1~\mathrm{eV}$, respectively. The DT cross-check yields $0.7\,\sigma$, $1.9\,\sigma$, and $3.1\,\sigma$, respectively. These estimates assume that the $T_{4c}$ signal and the non-resonant $e^+e^-\to\eta_cJ/\psi$ continuum contribute incoherently. Possible interference between the two amplitudes is not included in the present study and should be evaluated in future analyses.

In summary, the STCF, benefiting from its projected high luminosity and tunable beam energy, presents a promising platform to systematically search for and investigate fully-charmed tetraquark states near their production thresholds. Future experimental observations of the $\eta_{c}J/\psi$ channel are anticipated to yield valuable insights into the internal dynamics of compact tetraquarks and contribute to a deeper understanding of non-perturbative QCD.

\section*{Acknowledgements}
This work is supported by the National Key R\&D Program of China under Contracts No. 2022YFA1602200 and No. 2023YFA1607200; the National
	Natural Science Foundation of China (NSFC) under Contracts No. 12341501, No. 12341503, No. 12341504, and No. 12475091; the international partnership program of the Chinese Academy of Sciences Grant No. 211134\allowbreak{}KYSB20200057; Guangzhou Navigation Project No. 2024A04J6334. We thank the Hefei Comprehensive National Science Center for their strong support on the STCF key technology research project. We also thank the conveners of the STCF physics and software activities for their coordination and support, as well as the STCF software group for the development and maintenance of the simulation, reconstruction, and analysis software used in this work.

\bibliographystyle{cpl}
	\bibliography{refrence}

%Control: production of eprint (1) enabled
\begin{thebibliography}{30}%
\makeatletter
\providecommand \@ifxundefined [1]{%
 \@ifx{#1\undefined}
}%
\providecommand \@ifnum [1]{%
 \ifnum #1\expandafter \@firstoftwo
 \else \expandafter \@secondoftwo
 \fi
}%
\providecommand \@ifx [1]{%
 \ifx #1\expandafter \@firstoftwo
 \else \expandafter \@secondoftwo
 \fi
}%
\providecommand \natexlab [1]{#1}%
\providecommand \enquote  [1]{``#1''}%
\providecommand \bibnamefont  [1]{#1}%
\providecommand \bibfnamefont [1]{#1}%
\providecommand \citenamefont [1]{#1}%
\providecommand \href@noop [0]{\@secondoftwo}%
\providecommand \href [0]{\begingroup \@sanitize@url \@href}%
\providecommand \@href[1]{\@@startlink{#1}\@@href}%
\providecommand \@@href[1]{\endgroup#1\@@endlink}%
\providecommand \@sanitize@url [0]{\catcode `\\12\catcode `\$12\catcode
  `\&12\catcode `\#12\catcode `\^12\catcode `\_12\catcode `\%12\relax}%
\providecommand \@@startlink[1]{}%
\providecommand \@@endlink[0]{}%
\providecommand \url  [0]{\begingroup\@sanitize@url \@url }%
\providecommand \@url [1]{\endgroup\@href {#1}{\urlprefix }}%
\providecommand \urlprefix  [0]{URL }%
\providecommand \Eprint [0]{\href }%
\providecommand \doibase [0]{https://doi.org/}%
\providecommand \selectlanguage [0]{\@gobble}%
\providecommand \bibinfo  [0]{\@secondoftwo}%
\providecommand \bibfield  [0]{\@secondoftwo}%
\providecommand \translation [1]{[#1]}%
\providecommand \BibitemOpen [0]{}%
\providecommand \bibitemStop [0]{}%
\providecommand \bibitemNoStop [0]{.\EOS\space}%
\providecommand \EOS [0]{\spacefactor3000\relax}%
\providecommand \BibitemShut  [1]{\csname bibitem#1\endcsname}%
\let\auto@bib@innerbib\@empty
%</preamble>
\bibitem {PhysRevLett.91.262001}%
  \BibitemOpen
  \bibfield  {author} {\bibinfo {author} {\bibfnamefont {S.~K.}\ \bibnamefont
  {Choi}} \emph {et~al.} (\bibinfo {collaboration} {Belle Collaboration}),\
  }\href {https://doi.org/10.1103/PhysRevLett.91.262001} {\bibfield  {journal}
  {\bibinfo  {journal} {Phys. Rev. Lett.}\ }\textbf {\bibinfo {volume} {91}},\
  \bibinfo {pages} {262001} (\bibinfo {year} {2003})}\BibitemShut {NoStop}%
\bibitem {PhysRevLett.110.252001}%
  \BibitemOpen
  \bibfield  {author} {\bibinfo {author} {\bibfnamefont {M.}~\bibnamefont
  {Ablikim}} \emph {et~al.} (\bibinfo {collaboration} {BESIII Collaboration}),\
  }\href {https://doi.org/10.1103/PhysRevLett.110.252001} {\bibfield  {journal}
  {\bibinfo  {journal} {Phys. Rev. Lett.}\ }\textbf {\bibinfo {volume} {110}},\
  \bibinfo {pages} {252001} (\bibinfo {year} {2013})}\BibitemShut {NoStop}%
\bibitem {PhysRevLett.110.252002}%
  \BibitemOpen
  \bibfield  {author} {\bibinfo {author} {\bibfnamefont {Z.~Q.}\ \bibnamefont
  {Liu}} \emph {et~al.} (\bibinfo {collaboration} {Belle Collaboration}),\
  }\href {https://doi.org/10.1103/PhysRevLett.110.252002} {\bibfield  {journal}
  {\bibinfo  {journal} {Phys. Rev. Lett.}\ }\textbf {\bibinfo {volume} {110}},\
  \bibinfo {pages} {252002} (\bibinfo {year} {2013})}\BibitemShut {NoStop}%
\bibitem {PhysRevLett.115.072001}%
  \BibitemOpen
  \bibfield  {author} {\bibinfo {author} {\bibfnamefont {R.}~\bibnamefont
  {Aaij}} \emph {et~al.} (\bibinfo {collaboration} {LHCb Collaboration}),\
  }\href {https://doi.org/10.1103/PhysRevLett.115.072001} {\bibfield  {journal}
  {\bibinfo  {journal} {Phys. Rev. Lett.}\ }\textbf {\bibinfo {volume} {115}},\
  \bibinfo {pages} {072001} (\bibinfo {year} {2015})}\BibitemShut {NoStop}%
\bibitem {lhcbcollaborationObservationStructureJ2020a}%
  \BibitemOpen
  \bibfield  {author} {\bibinfo {author} {\bibfnamefont {R.}~\bibnamefont
  {Aaij}} \emph {et~al.} (\bibinfo {collaboration} {LHCb Collaboration}),\
  }\href {https://doi.org/10.1016/j.scib.2020.08.032} {\bibfield  {journal}
  {\bibinfo  {journal} {Sci. Bull.}\ }\textbf {\bibinfo {volume} {65}},\
  \bibinfo {pages} {1983} (\bibinfo {year} {2020})}\BibitemShut {NoStop}%
\bibitem {ATLAS2023bft}%
  \BibitemOpen
  \bibfield  {author} {\bibinfo {author} {\bibfnamefont {G.}~\bibnamefont
  {Aad}} \emph {et~al.} (\bibinfo {collaboration} {ATLAS Collaboration}),\
  }\href {https://doi.org/10.1103/PhysRevLett.131.151902} {\bibfield  {journal}
  {\bibinfo  {journal} {Phys. Rev. Lett.}\ }\textbf {\bibinfo {volume} {131}},\
  \bibinfo {pages} {151902} (\bibinfo {year} {2023})}\BibitemShut {NoStop}%
\bibitem {CMS2023owd}%
  \BibitemOpen
  \bibfield  {author} {\bibinfo {author} {\bibfnamefont {A.}~\bibnamefont
  {Hayrapetyan}} \emph {et~al.} (\bibinfo {collaboration} {CMS
  Collaboration}),\ }\href {https://doi.org/10.1103/PhysRevLett.132.111901}
  {\bibfield  {journal} {\bibinfo  {journal} {Phys. Rev. Lett.}\ }\textbf
  {\bibinfo {volume} {132}},\ \bibinfo {pages} {111901} (\bibinfo {year}
  {2024})}\BibitemShut {NoStop}%
\bibitem {CMS2025fpt}%
  \BibitemOpen
  \bibfield  {author} {\bibinfo {author} {\bibfnamefont {A.}~\bibnamefont
  {Hayrapetyan}} \emph {et~al.} (\bibinfo {collaboration} {CMS
  Collaboration}),\ }\href {https://doi.org/10.1038/s41586-025-09711-7}
  {\bibfield  {journal} {\bibinfo  {journal} {Nature}\ }\textbf {\bibinfo
  {volume} {648}},\ \bibinfo {pages} {58} (\bibinfo {year} {2025})}\BibitemShut
  {NoStop}%
\bibitem {chenPwaveFullyCharm2024}%
  \BibitemOpen
  \bibfield  {author} {\bibinfo {author} {\bibfnamefont {Z.-Z.}\ \bibnamefont
  {Chen}}, \bibinfo {author} {\bibfnamefont {X.-L.}\ \bibnamefont {Chen}},
  \bibinfo {author} {\bibfnamefont {P.-F.}\ \bibnamefont {Yang}},\ and\
  \bibinfo {author} {\bibfnamefont {W.}~\bibnamefont {Chen}},\ }\href
  {https://doi.org/10.1103/PhysRevD.109.094011} {\bibfield  {journal} {\bibinfo
   {journal} {Phys. Rev. D}\ }\textbf {\bibinfo {volume} {109}},\ \bibinfo
  {pages} {094011} (\bibinfo {year} {2024})}\BibitemShut {NoStop}%
\bibitem {faustovFullyheavyTetraquarkSpectroscopy2022}%
  \BibitemOpen
  \bibfield  {author} {\bibinfo {author} {\bibfnamefont {R.~N.}\ \bibnamefont
  {Faustov}}, \bibinfo {author} {\bibfnamefont {V.~O.}\ \bibnamefont
  {Galkin}},\ and\ \bibinfo {author} {\bibfnamefont {E.~M.}\ \bibnamefont
  {Savchenko}},\ }\href {https://doi.org/10.3390/sym14122504} {\bibfield
  {journal} {\bibinfo  {journal} {Symmetry}\ }\textbf {\bibinfo {volume}
  {14}},\ \bibinfo {pages} {2504} (\bibinfo {year} {2022})}\BibitemShut
  {NoStop}%
\bibitem {Dong2022sef}%
  \BibitemOpen
  \bibfield  {author} {\bibinfo {author} {\bibfnamefont {W.-C.}\ \bibnamefont
  {Dong}}\ and\ \bibinfo {author} {\bibfnamefont {Z.-G.}\ \bibnamefont
  {Wang}},\ }\href {https://doi.org/10.1103/PhysRevD.107.074010} {\bibfield
  {journal} {\bibinfo  {journal} {Phys. Rev. D}\ }\textbf {\bibinfo {volume}
  {107}},\ \bibinfo {pages} {074010} (\bibinfo {year} {2023})}\BibitemShut
  {NoStop}%
\bibitem {debastianiNonrelativisticModel$ccbarcbarc$2019}%
  \BibitemOpen
  \bibfield  {author} {\bibinfo {author} {\bibfnamefont {V.~R.}\ \bibnamefont
  {Debastiani}}\ and\ \bibinfo {author} {\bibfnamefont {F.~S.}\ \bibnamefont
  {Navarra}},\ }\href {https://doi.org/10.1088/1674-1137/43/1/013105}
  {\bibfield  {journal} {\bibinfo  {journal} {Chin. Phys. C}\ }\textbf
  {\bibinfo {volume} {43}},\ \bibinfo {pages} {013105} (\bibinfo {year}
  {2019})}\BibitemShut {NoStop}%
\bibitem {liu2020eha}%
  \BibitemOpen
  \bibfield  {author} {\bibinfo {author} {\bibfnamefont {M.-S.}\ \bibnamefont
  {Liu}}, \bibinfo {author} {\bibfnamefont {F.-X.}\ \bibnamefont {Liu}},
  \bibinfo {author} {\bibfnamefont {X.-H.}\ \bibnamefont {Zhong}},\ and\
  \bibinfo {author} {\bibfnamefont {Q.}~\bibnamefont {Zhao}},\ }\href
  {https://doi.org/10.1103/PhysRevD.109.076017} {\bibfield  {journal} {\bibinfo
   {journal} {Phys. Rev. D}\ }\textbf {\bibinfo {volume} {109}},\ \bibinfo
  {pages} {076017} (\bibinfo {year} {2024})}\BibitemShut {NoStop}%
\bibitem {wangHigherFullycharmedTetraquarks2021}%
  \BibitemOpen
  \bibfield  {author} {\bibinfo {author} {\bibfnamefont {G.-J.}\ \bibnamefont
  {Wang}}, \bibinfo {author} {\bibfnamefont {L.}~\bibnamefont {Meng}}, \bibinfo
  {author} {\bibfnamefont {M.}~\bibnamefont {Oka}},\ and\ \bibinfo {author}
  {\bibfnamefont {S.-L.}\ \bibnamefont {Zhu}},\ }\href
  {https://doi.org/10.1103/PhysRevD.104.036016} {\bibfield  {journal} {\bibinfo
   {journal} {Phys. Rev. D}\ }\textbf {\bibinfo {volume} {104}},\ \bibinfo
  {pages} {036016} (\bibinfo {year} {2021})}\BibitemShut {NoStop}%
\bibitem {yinSearchDoublecharmoniumState2023}%
  \BibitemOpen
  \bibfield  {author} {\bibinfo {author} {\bibfnamefont {J.~H.}\ \bibnamefont
  {Yin}} \emph {et~al.} (\bibinfo {collaboration} {Belle Collaboration}),\
  }\href {https://doi.org/10.1007/JHEP08(2023)121} {\bibfield  {journal}
  {\bibinfo  {journal} {JHEP}\ }\textbf {\bibinfo {volume} {08}},\ \bibinfo
  {pages} {121} (\bibinfo {year} {2023})}\BibitemShut {NoStop}%
\bibitem {achasovSTCFConceptualDesign2023}%
  \BibitemOpen
  \bibfield  {author} {\bibinfo {author} {\bibfnamefont {M.}~\bibnamefont
  {Achasov}} \emph {et~al.},\ }\href
  {https://doi.org/10.1007/s11467-023-1333-z} {\bibfield  {journal} {\bibinfo
  {journal} {Front. Phys. (Beijing)}\ }\textbf {\bibinfo {volume} {19}},\
  \bibinfo {pages} {14701} (\bibinfo {year} {2024})}\BibitemShut {NoStop}%
\bibitem {Lyu:2021tlb}%
  \BibitemOpen
  \bibfield  {author} {\bibinfo {author} {\bibfnamefont {X.-R.}\ \bibnamefont
  {Lyu}} (\bibinfo {collaboration} {STCF Working Group}),\ }\href
  {https://doi.org/10.22323/1.391.0060} {\bibfield  {journal} {\bibinfo
  {journal} {PoS}\ }\textbf {\bibinfo {volume} {BEAUTY2020}},\ \bibinfo {pages}
  {060} (\bibinfo {year} {2021})}\BibitemShut {NoStop}%
\bibitem {Ai:2024yqx}%
  \BibitemOpen
  \bibfield  {author} {\bibinfo {author} {\bibfnamefont {X.}~\bibnamefont
  {Ai}}, \bibinfo {author} {\bibfnamefont {X.}~\bibnamefont {Huang}}, \bibinfo
  {author} {\bibfnamefont {T.}~\bibnamefont {Li}}, \bibinfo {author}
  {\bibfnamefont {B.}~\bibnamefont {Qi}},\ and\ \bibinfo {author}
  {\bibfnamefont {X.}~\bibnamefont {Qin}},\ }\href
  {https://doi.org/10.1142/S0217732324400066} {\bibfield  {journal} {\bibinfo
  {journal} {Mod. Phys. Lett. A}\ }\textbf {\bibinfo {volume} {39}},\ \bibinfo
  {pages} {2440006} (\bibinfo {year} {2024})}\BibitemShut {NoStop}%
\bibitem {Li:2024tuy}%
  \BibitemOpen
  \bibfield  {author} {\bibinfo {author} {\bibfnamefont {T.}~\bibnamefont
  {Li}}, \bibinfo {author} {\bibfnamefont {W.}~\bibnamefont {Huang}}, \bibinfo
  {author} {\bibfnamefont {X.}~\bibnamefont {Huang}}, \bibinfo {author}
  {\bibfnamefont {X.}~\bibnamefont {Ai}}, \bibinfo {author} {\bibfnamefont
  {H.}~\bibnamefont {Li}},\ and\ \bibinfo {author} {\bibfnamefont
  {D.}~\bibnamefont {Liu}},\ }\href
  {https://doi.org/10.1051/epjconf/202429503025} {\bibfield  {journal}
  {\bibinfo  {journal} {EPJ Web Conf.}\ }\textbf {\bibinfo {volume} {295}},\
  \bibinfo {pages} {03025} (\bibinfo {year} {2024}{\natexlab{a}})}\BibitemShut
  {NoStop}%
\bibitem {Li:2024isl}%
  \BibitemOpen
  \bibfield  {author} {\bibinfo {author} {\bibfnamefont {T.}~\bibnamefont
  {Li}}, \bibinfo {author} {\bibfnamefont {X.}~\bibnamefont {Huang}}, \bibinfo
  {author} {\bibfnamefont {W.}~\bibnamefont {Huang}}, \bibinfo {author}
  {\bibfnamefont {X.}~\bibnamefont {Qin}}, \bibinfo {author} {\bibfnamefont
  {X.}~\bibnamefont {Ai}},\ and\ \bibinfo {author} {\bibfnamefont
  {B.}~\bibnamefont {Qi}},\ }\href {https://doi.org/10.1142/S0217732324400121}
  {\bibfield  {journal} {\bibinfo  {journal} {Mod. Phys. Lett. A}\ }\textbf
  {\bibinfo {volume} {39}},\ \bibinfo {pages} {2440012} (\bibinfo {year}
  {2024}{\natexlab{b}})}\BibitemShut {NoStop}%
\bibitem {Huang:2023kog}%
  \BibitemOpen
  \bibfield  {author} {\bibinfo {author} {\bibfnamefont {W.~H.}\ \bibnamefont
  {Huang}}, \bibinfo {author} {\bibfnamefont {T.}~\bibnamefont {Li}}, \bibinfo
  {author} {\bibfnamefont {Q.~Y.}\ \bibnamefont {Li}}, \bibinfo {author}
  {\bibfnamefont {H.}~\bibnamefont {Li}}, \bibinfo {author} {\bibfnamefont
  {D.}~\bibnamefont {Liu}},\ and\ \bibinfo {author} {\bibfnamefont {X.~T.}\
  \bibnamefont {Huang}},\ }\href
  {https://doi.org/10.1088/1742-6596/2438/1/012054} {\bibfield  {journal}
  {\bibinfo  {journal} {J. Phys. Conf. Ser.}\ }\textbf {\bibinfo {volume}
  {2438}},\ \bibinfo {pages} {012054} (\bibinfo {year} {2023})}\BibitemShut
  {NoStop}%
\bibitem {Ivanchenko:2003xp}%
  \BibitemOpen
  \bibfield  {author} {\bibinfo {author} {\bibfnamefont {V.~N.}\ \bibnamefont
  {Ivanchenko}} (\bibinfo {collaboration} {Geant4 Collaboration}),\ }\href
  {https://doi.org/10.1016/S0168-9002(03)00538-2} {\bibfield  {journal}
  {\bibinfo  {journal} {Nucl. Instrum. Meth. A}\ }\textbf {\bibinfo {volume}
  {502}},\ \bibinfo {pages} {666} (\bibinfo {year} {2003})}\BibitemShut
  {NoStop}%
\bibitem {Shi:2025qmb}%
  \BibitemOpen
  \bibfield  {author} {\bibinfo {author} {\bibfnamefont {Q.}~\bibnamefont
  {Shi}}, \bibinfo {author} {\bibfnamefont {T.}~\bibnamefont {Li}},\ and\
  \bibinfo {author} {\bibfnamefont {X.}~\bibnamefont {Huang}},\ }\href
  {https://doi.org/10.1007/s41781-025-00131-w} {\bibfield  {journal} {\bibinfo
  {journal} {Comput. Softw. Big Sci.}\ }\textbf {\bibinfo {volume} {9}},\
  \bibinfo {pages} {3} (\bibinfo {year} {2025})}\BibitemShut {NoStop}%
\bibitem {Zhou:2024tio}%
  \BibitemOpen
  \bibfield  {author} {\bibinfo {author} {\bibfnamefont {H.}~\bibnamefont
  {Zhou}}, \bibinfo {author} {\bibfnamefont {K.}~\bibnamefont {Sun}}, \bibinfo
  {author} {\bibfnamefont {Z.}~\bibnamefont {Lu}}, \bibinfo {author}
  {\bibfnamefont {H.}~\bibnamefont {Li}}, \bibinfo {author} {\bibfnamefont
  {X.}~\bibnamefont {Ai}}, \bibinfo {author} {\bibfnamefont {J.}~\bibnamefont
  {Zhang}}, \bibinfo {author} {\bibfnamefont {X.}~\bibnamefont {Huang}},\ and\
  \bibinfo {author} {\bibfnamefont {J.}~\bibnamefont {Liu}},\ }\href
  {https://doi.org/10.1016/j.nima.2025.170357} {\bibfield  {journal} {\bibinfo
  {journal} {Nucl. Instrum. Meth. A}\ }\textbf {\bibinfo {volume} {1075}},\
  \bibinfo {pages} {170357} (\bibinfo {year} {2025})}\BibitemShut {NoStop}%
\bibitem {Jia:2025ufk}%
  \BibitemOpen
  \bibfield  {author} {\bibinfo {author} {\bibfnamefont {X.}~\bibnamefont
  {Jia}}, \bibinfo {author} {\bibfnamefont {X.}~\bibnamefont {Qin}}, \bibinfo
  {author} {\bibfnamefont {T.}~\bibnamefont {Li}}, \bibinfo {author}
  {\bibfnamefont {X.}~\bibnamefont {Zhang}}, \bibinfo {author} {\bibfnamefont
  {X.}~\bibnamefont {Hu}}, \bibinfo {author} {\bibfnamefont {S.}~\bibnamefont
  {Song}}, \bibinfo {author} {\bibfnamefont {H.}~\bibnamefont {Zhou}}, \bibinfo
  {author} {\bibfnamefont {X.}~\bibnamefont {Ai}}, \bibinfo {author}
  {\bibfnamefont {J.}~\bibnamefont {Zhang}},\ and\ \bibinfo {author}
  {\bibfnamefont {X.}~\bibnamefont {Huang}},\ }\href
  {https://doi.org/10.1088/1748-0221/21/03/P03009} {\bibfield  {journal}
  {\bibinfo  {journal} {JINST}\ }\textbf {\bibinfo {volume} {21}}\bibfield
  {number} {\bibinfo  {number} { (03)},\ \bibinfo {pages} {P03009}} (\bibinfo
  {year} {2026})}\BibitemShut {NoStop}%
\bibitem {Zhai:2025qke}%
  \BibitemOpen
  \bibfield  {author} {\bibinfo {author} {\bibfnamefont {Y.}~\bibnamefont
  {Zhai}}\ and\ \bibinfo {author} {\bibfnamefont {Z.}~\bibnamefont {Yao}},\
  }\href {https://doi.org/10.22323/1.476.0847} {\bibfield  {journal} {\bibinfo
  {journal} {PoS}\ }\textbf {\bibinfo {volume} {ICHEP2024}},\ \bibinfo {pages}
  {847} (\bibinfo {year} {2025})}\BibitemShut {NoStop}%
\bibitem {Richter-Was:1992hxq}%
  \BibitemOpen
  \bibfield  {author} {\bibinfo {author} {\bibfnamefont {E.}~\bibnamefont
  {Richter-Was}},\ }\href {https://doi.org/10.1016/0370-2693(93)90062-M}
  {\bibfield  {journal} {\bibinfo  {journal} {Phys. Lett. B}\ }\textbf
  {\bibinfo {volume} {303}},\ \bibinfo {pages} {163} (\bibinfo {year}
  {1993})}\BibitemShut {NoStop}%
\bibitem {wang2009resonanceparametermeasurementluminosity}%
  \BibitemOpen
  \bibfield  {author} {\bibinfo {author} {\bibfnamefont {P.}~\bibnamefont
  {Wang}}, \bibinfo {author} {\bibfnamefont {Y.~S.}\ \bibnamefont {Zhu}},\ and\
  \bibinfo {author} {\bibfnamefont {X.~H.}\ \bibnamefont {Mo}},\ }\href
  {https://arxiv.org/abs/0907.0734} {\bibinfo {title} {On resonance parameter
  measurement and luminosity determination at $e^+e^-$ collider}} (\bibinfo
  {year} {2009})\BibitemShut {NoStop}%
\bibitem {QQQQ2016}%
  \BibitemOpen
  \bibfield  {author} {\bibinfo {author} {\bibfnamefont {M.}~\bibnamefont
  {Karliner}}, \bibinfo {author} {\bibfnamefont {S.}~\bibnamefont {Nussinov}},\
  and\ \bibinfo {author} {\bibfnamefont {J.~L.}\ \bibnamefont {Rosner}},\
  }\href {https://doi.org/10.1103/PhysRevD.95.034011} {\bibfield  {journal}
  {\bibinfo  {journal} {Phys. Rev. D}\ }\textbf {\bibinfo {volume} {95}},\
  \bibinfo {pages} {034011} (\bibinfo {year} {2017})}\BibitemShut {NoStop}%
\bibitem {BelleII2026heb}%
  \BibitemOpen
  \bibfield  {author} {\bibinfo {author} {\bibfnamefont {M.}~\bibnamefont
  {Abumusabh}} \emph {et~al.} (\bibinfo {collaboration} {Belle-II
  Collaboration}),\ }\href {https://doi.org/10.1103/f8qf-b284} {\bibfield
  {journal} {\bibinfo  {journal} {Phys. Rev. D}\ }\textbf {\bibinfo {volume}
  {113}},\ \bibinfo {pages} {112010} (\bibinfo {year} {2026})}\BibitemShut
  {NoStop}%
\end{thebibliography}%
% ======================================================================
% Supplementary Material (copied into this arXiv source)
% ======================================================================
\clearpage
\onecolumn
\renewcommand{\tablename}{Table}
\renewcommand{\figurename}{Figure}
\newcommand{\supptablefont}{\small}

\sisetup{
	separate-uncertainty = true,
	table-number-alignment = center
}

	\begin{center}
		{\Large Supplementary Material}\\[2mm]
{\large Feasibility Study of $e^+e^-\to \eta_cJ/\psi$ Production and Fully-Charmed Tetraquark Searches at STCF}
	\end{center}
	
\section*{S1. ST and DT Efficiencies and Correction Factors}

The ST detection efficiencies are determined independently at each energy point using 50,000 simulated signal events for each $J/\psi$ decay channel. Table~\ref{tab:supp_st_inputs} lists the ST efficiencies and the correction factors used in the ST cross-section calculation. Here, $f_{\rm ISR}$ and $f_{\rm vac}$ denote the initial-state-radiation and vacuum-polarization correction factors, respectively, and $f_{\rm corr}=f_{\rm ISR}f_{\rm vac}$. Final-state radiation is included in the simulation with \textsc{Photos} and is not assigned a separate correction factor. The DT efficiencies are listed separately in Table~\ref{tab:supp_dt_efficiencies}.

\begin{table}[!htbp]
	\centering
	\caption{ST efficiencies and correction factors at the scan points.}
	\label{tab:supp_st_inputs}
	\supptablefont
\begin{tabular*}{0.95\textwidth}{@{\extracolsep{\fill}}cccccc@{}}
			\toprule
			$\sqrt{s}$ (MeV) & $\varepsilon_{ee}$ (\%) & $\varepsilon_{\mu\mu}$ (\%) & $f_{\rm ISR}$ ($10^{-2}$) & $f_{\rm vac}$ ($10^{-2}$) & $f_{\rm corr}$ ($10^{-2}$) \\
			\midrule
			6710 & 65.0 & 79.7 & 65.4 & 106.5 & 69.6 \\
			6715 & 64.8 & 79.5 & 67.7 & 106.5 & 72.1 \\
			6720 & 64.9 & 79.8 & 69.4 & 106.5 & 73.9 \\
			6725 & 65.1 & 79.4 & 70.7 & 106.5 & 75.2 \\
			6730 & 64.7 & 79.6 & 71.7 & 106.5 & 76.4 \\
			6735 & 65.0 & 79.1 & 72.6 & 106.5 & 77.3 \\
			6740 & 64.8 & 79.4 & 73.3 & 106.5 & 78.0 \\
			6745 & 65.2 & 79.6 & 73.6 & 106.5 & 78.4 \\
			6750 & 64.9 & 79.5 & 73.9 & 106.5 & 78.7 \\
			6755 & 65.1 & 79.3 & 75.4 & 106.5 & 80.2 \\
			6760 & 64.7 & 79.9 & 76.4 & 106.5 & 81.3 \\
			6765 & 65.0 & 79.4 & 77.0 & 106.5 & 82.0 \\
			6770 & 64.9 & 79.6 & 77.4 & 106.5 & 82.4 \\
			6775 & 64.7 & 79.5 & 77.8 & 106.5 & 82.9 \\
			6780 & 65.0 & 79.3 & 78.2 & 106.5 & 83.3 \\
			6785 & 64.8 & 79.7 & 78.6 & 106.5 & 83.7 \\
			6790 & 64.9 & 79.5 & 79.0 & 106.5 & 84.1 \\
			\bottomrule
\end{tabular*}
\end{table}

\begin{table}[!htbp]
	\centering
	\caption{DT detection efficiencies at the scan points.}
	\label{tab:supp_dt_efficiencies}
\supptablefont
	\begin{tabular*}{0.82\textwidth}{@{\extracolsep{\fill}}ccc@{}}
		\toprule
		$\sqrt{s}$ (MeV) & $\varepsilon_{ee}$ (\%) & $\varepsilon_{\mu\mu}$ (\%) \\
		\midrule
		6710 & 31.1 & 39.8 \\
		6715 & 31.0 & 40.2 \\
		6720 & 31.2 & 39.8 \\
		6725 & 30.9 & 40.3 \\
		6730 & 31.1 & 39.9 \\
		6735 & 30.9 & 40.2 \\
		6740 & 31.2 & 39.9 \\
		6745 & 31.0 & 40.3 \\
		6750 & 30.9 & 39.8 \\
		6755 & 31.2 & 40.3 \\
		6760 & 31.0 & 40.0 \\
		6765 & 31.2 & 40.1 \\
		6770 & 30.9 & 39.8 \\
		6775 & 31.1 & 40.3 \\
		6780 & 30.9 & 39.7 \\
		6785 & 31.2 & 40.4 \\
		6790 & 31.0 & 39.9 \\
		\bottomrule
	\end{tabular*}
\end{table}

\clearpage
\section*{S2. Cross-Section Inputs}

Table~\ref{tab:supp_cross_sections} lists the point-by-point Born cross sections used for the $T_{4c}$ signal and the non-resonant $e^+e^-\to\eta_cJ/\psi$ continuum. The signal cross section is proportional to $\Gamma_{ee}$; the three signal columns correspond to $\Gamma_{ee}=0.25$, $0.5$, and $1~\mathrm{eV}$, respectively. The total cross sections are obtained by adding the signal and continuum contributions.

\begin{table}[!htbp]
	\centering
	\caption{Signal and continuum Born cross sections at the scan points.}
	\label{tab:supp_cross_sections}
\supptablefont
	\begin{tabular*}{\textwidth}{@{\extracolsep{\fill}}cccccccc@{}}
			\toprule
			& \multicolumn{3}{c}{$\sigma_{\rm sig}$ (fb)} & $\sigma_{\rm cont}$ (fb) & \multicolumn{3}{c}{$\sigma_{\rm tot}$ (fb)} \\
			\cmidrule(lr){2-4}\cmidrule(lr){6-8}
			$\sqrt{s}$ (GeV) & $0.25~\mathrm{eV}$ & $0.5~\mathrm{eV}$ & $1~\mathrm{eV}$ & & $0.25~\mathrm{eV}$ & $0.5~\mathrm{eV}$ & $1~\mathrm{eV}$ \\
			\midrule
			6.710 & 2.1 & 4.2 & 8.4 & 754.8 & 756.9 & 759.0 & 763.2 \\
			6.715 & 2.7 & 5.4 & 10.9 & 752.9 & 755.6 & 758.3 & 763.7 \\
			6.720 & 3.6 & 7.3 & 14.6 & 750.9 & 754.5 & 758.2 & 765.5 \\
			6.725 & 5.1 & 10.3 & 20.6 & 748.9 & 754.1 & 759.2 & 769.5 \\
			6.730 & 7.8 & 15.5 & 31.0 & 746.9 & 754.7 & 762.4 & 777.9 \\
			6.735 & 12.8 & 25.7 & 51.3 & 744.9 & 757.7 & 770.6 & 796.2 \\
			6.740 & 24.1 & 48.2 & 96.3 & 742.9 & 767.0 & 791.0 & 839.2 \\
			6.745 & 50.9 & 101.8 & 203.6 & 740.8 & 791.7 & 842.6 & 944.4 \\
			6.750 & 81.0 & 162.0 & 324.0 & 738.8 & 819.8 & 900.8 & 1062.8 \\
			6.755 & 50.9 & 101.8 & 203.5 & 736.7 & 787.6 & 838.5 & 940.2 \\
			6.760 & 24.0 & 48.1 & 96.1 & 734.7 & 758.7 & 782.7 & 830.8 \\
			6.765 & 12.8 & 25.6 & 51.1 & 732.6 & 745.4 & 758.1 & 783.7 \\
			6.770 & 7.7 & 15.4 & 30.9 & 730.5 & 738.2 & 745.9 & 761.3 \\
			6.775 & 5.1 & 10.2 & 20.4 & 728.4 & 733.5 & 738.6 & 748.8 \\
			6.780 & 3.6 & 7.2 & 14.5 & 726.3 & 729.9 & 733.5 & 740.7 \\
			6.785 & 2.7 & 5.4 & 10.8 & 724.1 & 726.8 & 729.5 & 734.9 \\
			6.790 & 2.1 & 4.1 & 8.3 & 722.0 & 724.1 & 726.2 & 730.3 \\
			\bottomrule
	\end{tabular*}
\end{table}

\clearpage
\section*{S3. Fitted Single-Tag Signal Yields}

Tables~\ref{tab:supp_st_025}, \ref{tab:supp_st_05}, and \ref{tab:supp_st_1} summarize the fitted single-tag (ST) signal yields for the three $\Gamma_{ee}$ hypotheses. The quoted uncertainties are the fit errors at each energy point.

\begin{table}[H]
	\centering
	\caption{ST fitted signal yields for the $\Gamma_{ee}=0.25~\mathrm{eV}$ hypothesis.}
	\label{tab:supp_st_025}
\supptablefont
	\begin{tabular*}{0.95\textwidth}{@{\extracolsep{\fill}}ccccc@{}}
		\toprule
		$\sqrt{s}$ (MeV) & $N_{\rm sig}^{ee}$ & Err$_{ee}$ & $N_{\rm sig}^{\mu\mu}$ & Err$_{\mu\mu}$ \\
		\midrule
		6710 & 2268.6 & 48.5 & 2763.5 & 53.6 \\
		6715 & 2315.3 & 48.9 & 2732.5 & 53.2 \\
		6720 & 2323.8 & 46.9 & 2766.9 & 49.3 \\
		6725 & 2263.6 & 40.2 & 2780.4 & 53.7 \\
		6730 & 2329.7 & 49.1 & 2710.1 & 53.0 \\
		6735 & 2294.0 & 42.8 & 2727.4 & 53.3 \\
		6740 & 2363.3 & 49.6 & 2763.9 & 53.6 \\
		6745 & 2364.0 & 48.5 & 2858.2 & 54.5 \\
		6750 & 2531.3 & 53.1 & 3020.8 & 57.8 \\
		6755 & 2329.5 & 42.8 & 2830.1 & 41.5 \\
		6760 & 2260.6 & 38.2 & 2706.6 & 49.4 \\
		6765 & 2231.1 & 42.3 & 2688.8 & 31.3 \\
		6770 & 2233.2 & 47.9 & 2642.0 & 52.5 \\
		6775 & 2212.8 & 41.6 & 2668.0 & 52.7 \\
		6780 & 2176.7 & 33.3 & 2632.4 & 52.5 \\
		6785 & 2150.3 & 47.5 & 2559.4 & 53.5 \\
		6790 & 2197.7 & 49.8 & 2625.6 & 35.4 \\
		\bottomrule
	\end{tabular*}
\end{table}

\clearpage
\begin{table}[H]
	\centering
	\caption{ST fitted signal yields for the $\Gamma_{ee}=0.5~\mathrm{eV}$ hypothesis.}
	\label{tab:supp_st_05}
\supptablefont
	\begin{tabular*}{0.95\textwidth}{@{\extracolsep{\fill}}ccccc@{}}
		\toprule
		$\sqrt{s}$ (MeV) & $N_{\rm sig}^{ee}$ & Err$_{ee}$ & $N_{\rm sig}^{\mu\mu}$ & Err$_{\mu\mu}$ \\
		\midrule
		6710 & 2325.4 & 48.5 & 2826.8 & 55.7 \\
		6715 & 2379.7 & 50.0 & 2807.2 & 55.6 \\
		6720 & 2394.5 & 30.8 & 2828.3 & 55.8 \\
		6725 & 2311.7 & 49.5 & 2863.1 & 55.9 \\
		6730 & 2400.0 & 49.2 & 2804.0 & 55.0 \\
		6735 & 2372.6 & 50.8 & 2827.4 & 55.9 \\
		6740 & 2491.2 & 51.2 & 2914.1 & 55.5 \\
		6745 & 2575.5 & 53.1 & 3099.4 & 58.0 \\
		6750 & 2807.3 & 55.9 & 3331.5 & 60.6 \\
		6755 & 2527.7 & 52.8 & 3091.6 & 58.0 \\
		6760 & 2362.1 & 51.3 & 2848.9 & 55.9 \\
		6765 & 2321.9 & 50.7 & 2765.7 & 55.6 \\
		6770 & 2307.6 & 50.4 & 2716.6 & 54.9 \\
		6775 & 2261.0 & 49.8 & 2740.0 & 55.1 \\
		6780 & 2242.7 & 49.4 & 2666.9 & 54.7 \\
		6785 & 2176.9 & 49.6 & 2573.8 & 53.9 \\
		6790 & 2223.4 & 50.0 & 2689.0 & 55.0 \\
		\bottomrule
	\end{tabular*}
\end{table}

\clearpage
\begin{table}[H]
	\centering
	\caption{ST fitted signal yields for the $\Gamma_{ee}=1~\mathrm{eV}$ hypothesis.}
	\label{tab:supp_st_1}
\supptablefont
	\begin{tabular*}{0.95\textwidth}{@{\extracolsep{\fill}}ccccc@{}}
		\toprule
		$\sqrt{s}$ (MeV) & $N_{\rm sig}^{ee}$ & Err$_{ee}$ & $N_{\rm sig}^{\mu\mu}$ & Err$_{\mu\mu}$ \\
		\midrule
		6710 & 2334.2 & 44.5 & 2840.0 & 55.8 \\
		6715 & 2396.4 & 50.1 & 2823.3 & 55.8 \\
		6720 & 2420.2 & 37.3 & 2857.0 & 56.1 \\
		6725 & 2340.9 & 49.8 & 2897.3 & 56.2 \\
		6730 & 2445.1 & 49.7 & 2861.3 & 55.5 \\
		6735 & 2452.2 & 51.6 & 2926.1 & 56.7 \\
		6740 & 2649.6 & 52.7 & 3085.0 & 57.2 \\
		6745 & 2878.2 & 55.9 & 3500.8 & 62.0 \\
		6750 & 3315.5 & 60.3 & 3933.9 & 65.9 \\
		6755 & 2852.9 & 56.0 & 3452.8 & 61.0 \\
		6760 & 2510.9 & 52.8 & 3011.7 & 57.3 \\
		6765 & 2402.8 & 51.6 & 2847.8 & 56.4 \\
		6770 & 2357.4 & 51.0 & 2777.6 & 55.4 \\
		6775 & 2289.8 & 50.1 & 2782.7 & 55.5 \\
		6780 & 2261.1 & 49.6 & 2694.2 & 55.0 \\
		6785 & 2193.8 & 49.7 & 2592.8 & 54.1 \\
		6790 & 2238.9 & 50.1 & 2700.3 & 55.1 \\
		\bottomrule
	\end{tabular*}
\end{table}

\clearpage
\section*{S4. Fitted Double-Tag Signal Yields}

Tables~\ref{tab:supp_dt_025}, \ref{tab:supp_dt_05}, and \ref{tab:supp_dt_1} summarize the fitted double-tag (DT) signal yields for the three $\Gamma_{ee}$ hypotheses. The quoted uncertainties are the fit errors at each energy point.

\begin{table}[H]
	\centering
	\caption{DT fitted signal yields for the $\Gamma_{ee}=0.25~\mathrm{eV}$ hypothesis.}
	\label{tab:supp_dt_025}
\supptablefont
	\begin{tabular*}{0.95\textwidth}{@{\extracolsep{\fill}}ccccc@{}}
		\toprule
		$\sqrt{s}$ (MeV) & $N_{\rm sig}^{ee}$ & Err$_{ee}$ & $N_{\rm sig}^{\mu\mu}$ & Err$_{\mu\mu}$ \\
		\midrule
		6710 & 5.5 & 2.7 & 13.5 & 3.6 \\
		6715 & 14.6 & 3.7 & 13.0 & 3.4 \\
		6720 & 9.6 & 3.2 & 16.0 & 3.8 \\
		6725 & 17.1 & 4.3 & 13.0 & 3.4 \\
		6730 & 13.2 & 3.7 & 17.1 & 3.9 \\
		6735 & 11.5 & 3.3 & 15.1 & 3.7 \\
		6740 & 13.1 & 3.7 & 15.9 & 3.8 \\
		6745 & 12.0 & 3.2 & 13.0 & 3.4 \\
		6750 & 13.0 & 3.4 & 19.9 & 4.3 \\
		6755 & 9.5 & 3.3 & 17.0 & 3.9 \\
		6760 & 12.4 & 3.5 & 18.9 & 4.2 \\
		6765 & 16.0 & 3.8 & 15.0 & 3.7 \\
		6770 & 12.5 & 3.5 & 11.0 & 3.1 \\
		6775 & 6.0 & 2.3 & 11.0 & 3.1 \\
		6780 & 8.8 & 3.5 & 12.9 & 3.4 \\
		6785 & 6.9 & 2.5 & 6.9 & 2.3 \\
		6790 & 10.8 & 3.3 & 7.1 & 2.5 \\
		\bottomrule
	\end{tabular*}
\end{table}

\clearpage
\begin{table}[H]
	\centering
	\caption{DT fitted signal yields for the $\Gamma_{ee}=0.5~\mathrm{eV}$ hypothesis.}
	\label{tab:supp_dt_05}
\supptablefont
	\begin{tabular*}{0.95\textwidth}{@{\extracolsep{\fill}}ccccc@{}}
		\toprule
		$\sqrt{s}$ (MeV) & $N_{\rm sig}^{ee}$ & Err$_{ee}$ & $N_{\rm sig}^{\mu\mu}$ & Err$_{\mu\mu}$ \\
		\midrule
		6710 & 5.5 & 2.7 & 12.5 & 3.4 \\
		6715 & 14.6 & 3.7 & 13.0 & 3.4 \\
		6720 & 9.6 & 3.2 & 16.0 & 3.8 \\
		6725 & 18.1 & 4.1 & 13.0 & 3.4 \\
		6730 & 15.0 & 4.0 & 17.1 & 3.9 \\
		6735 & 13.5 & 3.6 & 14.1 & 3.6 \\
		6740 & 14.2 & 3.9 & 16.9 & 3.9 \\
		6745 & 14.0 & 3.5 & 14.0 & 3.5 \\
		6750 & 16.0 & 3.8 & 22.9 & 4.6 \\
		6755 & 10.6 & 3.5 & 18.0 & 4.0 \\
		6760 & 13.4 & 3.6 & 20.9 & 4.4 \\
		6765 & 16.0 & 3.8 & 15.0 & 3.7 \\
		6770 & 13.5 & 3.6 & 11.0 & 3.1 \\
		6775 & 6.0 & 2.3 & 12.9 & 3.4 \\
		6780 & 8.5 & 3.5 & 12.9 & 3.4 \\
		6785 & 6.9 & 2.5 & 9.0 & 2.7 \\
		6790 & 10.8 & 3.3 & 15.1 & 3.8 \\
		\bottomrule
	\end{tabular*}
\end{table}

\clearpage
\begin{table}[H]
	\centering
	\caption{DT fitted signal yields for the $\Gamma_{ee}=1~\mathrm{eV}$ hypothesis.}
	\label{tab:supp_dt_1}
\supptablefont
	\begin{tabular*}{0.95\textwidth}{@{\extracolsep{\fill}}ccccc@{}}
		\toprule
		$\sqrt{s}$ (MeV) & $N_{\rm sig}^{ee}$ & Err$_{ee}$ & $N_{\rm sig}^{\mu\mu}$ & Err$_{\mu\mu}$ \\
		\midrule
		6710 & 6.8 & 2.9 & 13.5 & 3.6 \\
		6715 & 15.6 & 3.8 & 13.0 & 3.4 \\
		6720 & 9.6 & 3.2 & 16.0 & 3.8 \\
		6725 & 18.1 & 4.1 & 14.0 & 3.5 \\
		6730 & 16.2 & 4.1 & 17.9 & 4.0 \\
		6735 & 13.5 & 3.6 & 15.1 & 3.7 \\
		6740 & 14.2 & 3.9 & 18.0 & 4.1 \\
		6745 & 16.0 & 3.8 & 18.0 & 4.0 \\
		6750 & 19.0 & 4.2 & 27.0 & 5.0 \\
		6755 & 13.0 & 3.8 & 20.0 & 4.3 \\
		6760 & 13.4 & 3.6 & 21.9 & 4.5 \\
		6765 & 17.0 & 3.9 & 16.0 & 3.8 \\
		6770 & 13.5 & 3.6 & 11.0 & 3.1 \\
		6775 & 6.0 & 2.3 & 12.9 & 3.4 \\
		6780 & 8.5 & 3.5 & 12.9 & 3.4 \\
		6785 & 7.9 & 2.7 & 10.0 & 2.9 \\
		6790 & 10.8 & 3.3 & 15.1 & 3.8 \\
		\bottomrule
	\end{tabular*}
\end{table}

\clearpage
\section*{S5. Double-Tag Cross-Section Fits}

\begin{figure}[!htbp]
	\centering
	\begin{overpic}[width=0.32\textwidth]{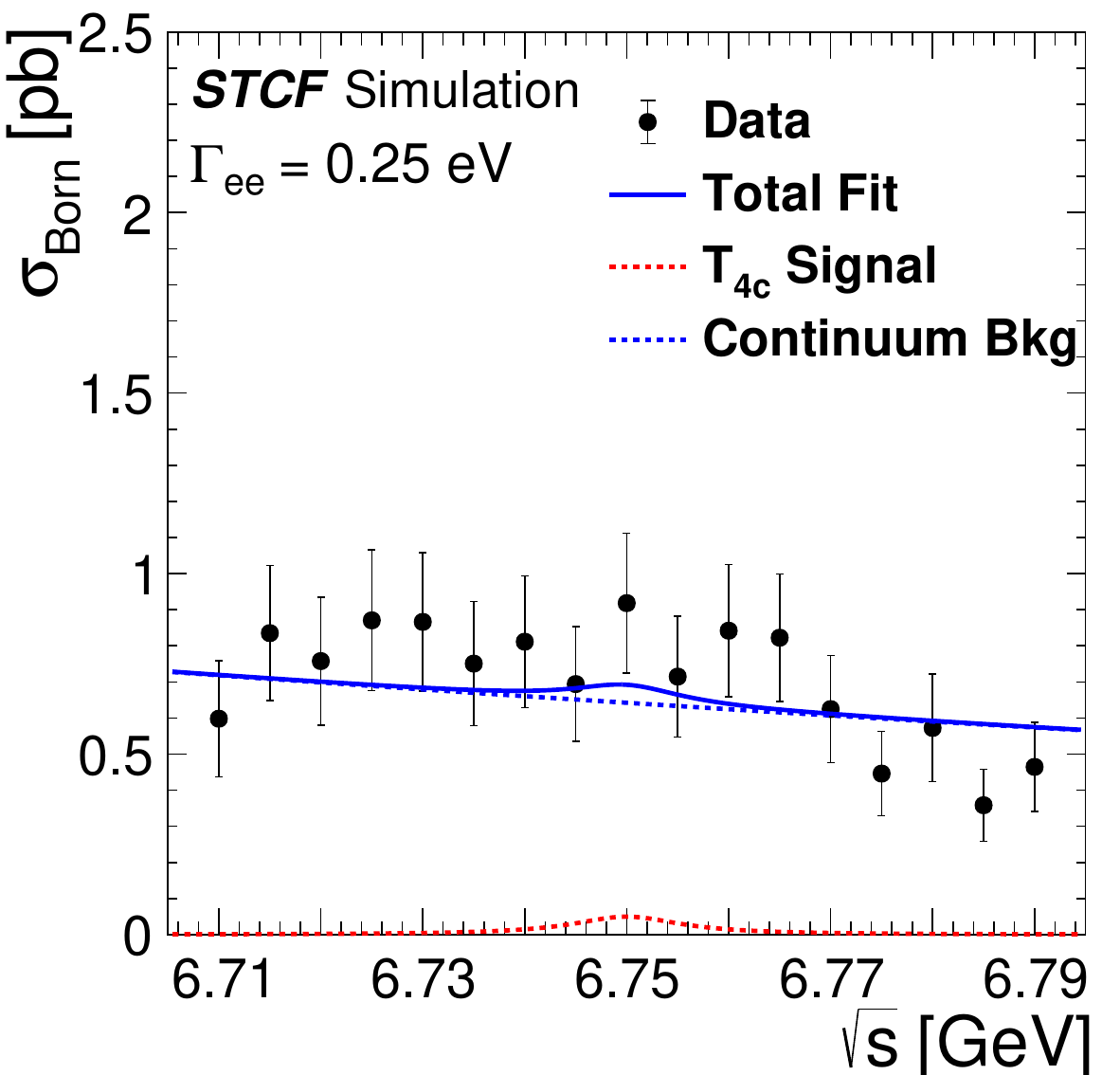}
		\put(80,18){(a)}       % 左上角
	\end{overpic}
	\hfill
	\begin{overpic}[width=0.32\textwidth]{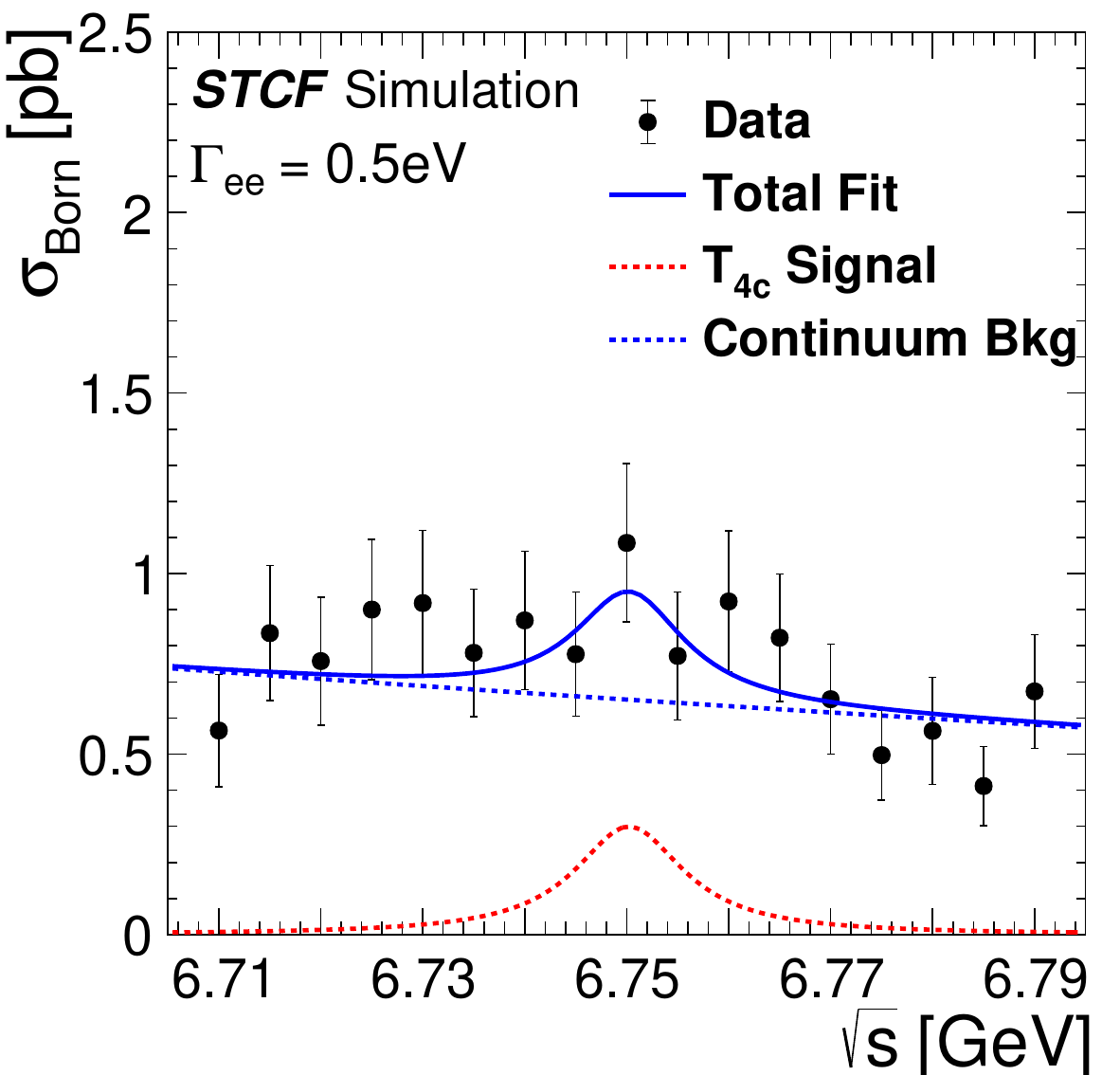}
		\put(80,18){(b)}
	\end{overpic}
	\hfill
	\begin{overpic}[width=0.32\textwidth]{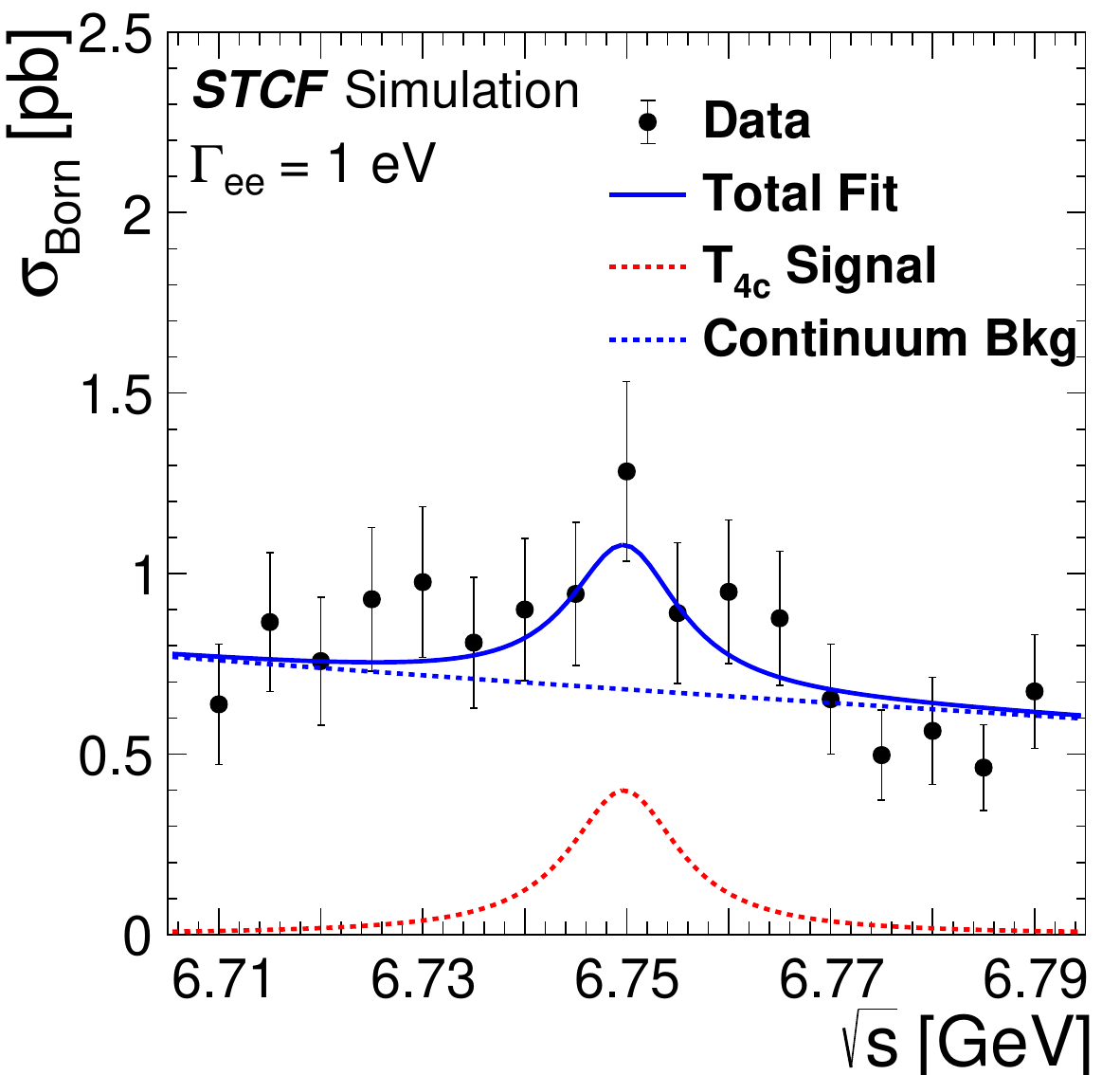}
		\put(80,18){(c)}
	\end{overpic}
	\caption{DT Born cross-section fits for (a) $\Gamma_{ee}=0.25~\mathrm{eV}$, ...}
	\label{fig:supp_dt_fits}
\end{figure}

Figure~\ref{fig:supp_dt_fits} shows the fitted double-tag (DT) Born cross-section distributions for the three $\Gamma_{ee}$ hypotheses. The solid blue curves represent the total fit, the dotted blue curves indicate the continuum component, and the dotted red curves show the $T_{4c}$ signal component.

\end{document}